\documentclass[aps,pre,twocolumn,superscriptaddress,natlib,showpacs]{revtex4-1}
\usepackage{natbib}
\usepackage{graphicx}
\usepackage{dcolumn}
\usepackage{bm}
\usepackage{color}
\usepackage{amssymb}
\usepackage{amsmath}
\usepackage{amscd}

\begin{document}

\title{Fractional electron transfer kinetics and a quantum breaking of ergodicity }
\author{Igor Goychuk}
\email{igoychuk@uni-potsdam.de}
\affiliation{Institute for Physics and Astronomy, University of Potsdam, 
Karl-Liebknecht-Str. 24/25, 14476 Potsdam-Golm, Germany}

\date{\today}

\begin{abstract}

The dissipative curve-crossing problem provides a paradigm for electron-transfer (ET) processes in condensed media. It establishes the simplest conceptual test bed to study the influence of the medium's dynamics on ET kinetics both on the ensemble level, and on the level of single particles. Single electron description is particularly important for nanoscaled systems like proteins, or molecular wires. Especially insightful is this framework in the semi-classical limit, where the environment can be treated classically, and an exact analytical treatment becomes feasible. Slow medium's dynamics is capable of enslaving ET and bringing it on the ensemble level from a quantum regime of non-adiabatic tunneling to the classical adiabatic regime, where electrons follow the nuclei rearrangements. This classical adiabatic textbook picture contradicts, however, in a very spectacular fashion to the statistics of single electron transitions, even in the Debye, memoryless media, also named Ohmic in the parlance of the famed spin-boson model. On the single particle level, ET always remains quantum, and this was named a quantum breaking of ergodicity in the adiabatic ET regime. What happens in the case of subdiffusive, fractional, or sub-Ohmic medium's dynamics, which is featured by power-law decaying dynamical memory effects typical, e.g. for protein macromolecules, and other viscoelastic media? Such a memory is vividly manifested by anomalous Cole-Cole dielectric response in such media. We address this question based both on accurate numerics and analytical theory. The ensemble theory remarkably agrees with the numerical dynamics of electronic populations, revealing a power law relaxation tail even in a profoundly non-adiabatic electron transfer regime. In other words, ET in such media should typically display  fractional kinetics. However, a profound difference with the numerically accurate results occurs for the distribution of residence times in the electronic states, both on the ensemble level and the level of single trajectories. Ergodicity is broken dynamically even in a more spectacular way than in the memoryless case. Our results question the applicability of all the existing and widely accepted ensemble theories of electron transfer in fractional, sub-Ohmic environments, on the level of single molecules, and provide a real challenge to face, both for theorists and experimentalists.


\end{abstract}


\maketitle

\section{Introduction}

Electron transfer (ET) is an important physical phenomenon across many research areas ranging from meso- and nanoscale physics, including physics of quantum dots and molecular electronics to molecular and chemical physics, as well biophysics \cite{Marcus56,Marcus57,Levich59,Hush58,Marcus60,Atkins,Nitzan, MayBook,Hush96,Barbara96,Fransuzov08,Efros,Bostick18}. It is central e.g. for bioenergetics \cite{Atkins,PhysicalBiology}. On nanoscale, the single-electron transfer is especially important and relevant, and the dynamics of the reaction coordinate coupled to ET generally should not be disregarded even for a long-range ET in proteins  and bioinspired ET reactions, as discussed in reviews \cite{GrayPNAS,Skourtis10} and references cited therein, e.g. \cite{Wei02,Kotelnikov02,Khoshtariya03,Liu05,Chakrabarti09}. Electron as a light particle is fundamentally quantum in its properties even if, e.g., the theory of adiabatic electron transport can be formulated as a purely classical theory on the ensemble level \cite{Marcus56, Marcus57, Marcus60, Nitzan, MayBook}. A common rationale behind this is that in such regime electrons follow adiabatically to the nuclear rearrangements, and nuclei can often be treated classically at sufficiently high temperatures \cite{Marcus56, Hush58, Atkins, Nitzan, MayBook}. This classical point of view has recently been challenged on the level of a single trajectory description by showing that the statistics of single-electron transitions between two diabatic quantum states is fundamentally different from the results expected from the classical theory of adiabatic ET \cite{TangJCP, PCCP17}. The reason for this lies in a fundamentally quantum nature of electron transitions, even in the adiabatic transport regime \cite{PCCP17}.

Hence, we are dealing with a truly quantum breaking of ergodicity in a seemingly classical adiabatic regime. This phenomenon should be distinguished from other non-ergodic effects caused by slow relaxation modes of the environment (classical breaking of ergodicity) \cite{GoychukPRE09, GoychukACP12}. In this respect, it is worth stressing that we mean here the genuine dynamical non-ergodic effects, entering through a \textit{relatively} slow dynamics of the one-dimensional reaction coordinate coupled to the electron transfer, rather than via a non-ergodic change of the medium's reorganization energy in polar solvents \cite{MatyushovACR, SeyediJPCL, Matyushov15}, or a local phase transition when the medium becomes temporally trapped in local minima of a multi-dimensional rough potential landscape \cite{Leite05}. The latter ones can also be important issues \textit{per se}. However, they are beyond the scope of this work dealing with a one-dimensional reaction coordinate description, as the simplest pertinent dynamical model \cite{Zusman, Garg, Tang96, Hartmann00, Casado03}. All the rest degrees of the environment are not coupled directly to the electron transfer dynamics and act as memory friction and the corresponding correlated thermal noise affecting the dynamics of the reaction coordinate. This physical picture leads ultimately to a dynamical breakdown of the rate description, like also in \cite{Leite05} within a very different model. The environment is assumed to be at thermal equilibrium, obeying the fluctuation-dissipation theorem (FDT) \cite{KuboBook, ZwanzigBook}. Here lies a fundamental difference with Refs. \cite{MatyushovACR,SeyediJPCL,Matyushov15}. However, our description can also be easily generalized to account for nonequilibrium noise or periodic field, produced either externally, or (noise) as a result of nonequilibrium conformational dynamics \cite{Goychuk94,PetrovPRE94,GoychukPRE95,GoychukPRE95,GoychukCPL96,GoychukPRE97,Hartmann00,GoychukPRL06,GoychukAP05}, as in various molecular machines \cite{Goychuk16}, including electron tunneling pumps \cite{GoychukMolSim06,GoychukAP05}.

One should also mention a general notion of ergodicity in the theory of stochastic processes \cite{Papoulis}. Namely, it concerns coincidence of single-trajectory time-averages over a very large (infinite in theory) time interval and the (infinite in theory) ensemble averages for the same process. The ergodicity in question can be defined and understood in various senses, e.g., in the mean value (most commonly used in statistical physics), in the variance of a random process, in the distribution of its assumed values, etc. \cite{Papoulis}. Whenever the ensemble and trajectory averages are different, we are speaking about broken ergodicity. In this work, we understand ergodicity in a kinetic sense \cite{Bologna02, PCCP17}. Namely, if an \textit{equilibrium} ensemble theory is capable of describing statistics of multiple subsequent single-electron transitions, revealed, e.g. in a very long ``on-off'' blinking recording \cite{Pelton, Efros},  the ergodicity in this kinetic sense holds, and, otherwise, it does not. Our primary focus in this paper is on what happens when quantum and classical dynamical breaking of ergodicity meet in anomalous, non-exponential ET transport kinetics, which cannot be rigorously characterized by a rate anymore.

In this respect, most experiments on ET in various molecules and molecular compounds are done on the ensemble level \cite{GrayPNAS,Wei02,Kotelnikov02,Khoshtariya03,Liu05,Chakrabarti09}. On the level of single molecules and quantum dots, they are less common and appeared first with the advance of single-molecular research \cite{Yang03, MinPRL, Pelton, Efros}. In Ref. \cite{Yang03}, for example, ET in proteins was studied both from the ensemble and single-molecular perspectives. ET kinetics in a naturally occurring flavin reductase (Fre)/flavin adenine dinucleotide (FAD) protein complex was fitted by sums of four exponentials in the case of ensemble kinetics and three exponentials in the case of single-molecular kinetics. The largest time constant of the ensemble measurements was $3.3$ ns, whereas for single-molecular measurements it was $2.26$ ns, with the largest transfer time measured of about $10$ ns. In this case, the ensemble and single-molecular measurements yielded similar results, which, however, should not be generally expected for ET in proteins. Furthermore, single-molecular measurements revealed long-time correlations in ET transfer times extending far beyond the mean ET transfer time \cite{Yang03}. The authors explained the origin of these correlations within a non-adiabatic ET theory with a time-modulated rate that follows a much slower conformational dynamics over a vast time range from milliseconds to tens of seconds. They used, in fact, a variant of the theory of dynamical rate disorder \cite{Zwanzig90, Agmon84, Sumi86, Gehlen94, Wang93, Wang95,LeeJCP03,GoychukJCP05,Wang08}, which is very different from the approach of this work, though it suits also well to describe intermittency and non-exponential kinetics of single-molecular reactions in fluctuating environments \cite{Wang93,Wang95,LeeJCP03,GoychukJCP05,Wang08}.  Conformational dynamics was described by a fractional diffusion in a parabolic potential, first \cite{Yang03} within a fractional Fokker-Planck equation approach \cite{MetzlerPRL99, Metzler00}, and later \cite{KouPRL} within a very different fractional Langevin equation approach \cite{LutzFLE, GoychukPRL07, GoychukPRE09, GoychukACP12}, which we also use in this work. A Mittag-Leffler relaxation law describes this dynamics \cite{MetzlerPRL99, Metzler00}, which corresponds to a Cole-Cole dielectric response \cite{GoychukPRE07Rapid, GoychukACP12}.

The theory of electron transfer is mostly based on such ensemble concepts as reduced density matrix and corresponding kinetic equations \cite{Marcus56, Marcus57, Marcus60, Nitzan, MayBook}. However, can we fully trust in such ensemble theory concepts and approaches when applied to single-molecular ET, especially in the case of non-Debye media featured by a fractional relaxation?
We address this challenging question within a dissipative curve-crossing problem \cite{Wolynes87,Nitzan,MayBook}, with the reaction coordinate treated classically. Here, the Zusman model of electron transfer generalized to non-Debye environments \cite{Tang96, TangMarcusPRL}, i.e., with the reaction coordinate coupled to a sub-Ohmic bosonic thermal bath \cite{WeissBook} instead of the standard Ohmic one \cite{Garg}, provides an ideal playground. This model corresponds to a subdiffusive motion of nuclei on the diabatic curve in a particular localized electronic state, which in the case of standard one-dimensional parabolic curves leads namely to a Mittag-Leffler relaxation of the reaction coordinate, when the inertial effects are neglected \cite{GoychukPRE07Rapid, GoychukACP12}. It is initially stretched exponential and then changes to a power law. Such a relaxation behavior corresponds precisely \cite{GoychukPRE07Rapid} to the Cole-Cole anomalous dielectric response \cite{ColeCole}, commonly measured in many molecular systems, including proteins, DNAs, cytosol of biological cells, and biological membranes \cite{Gabriel}.
Even bounded water in many biological tissues displays a Cole-Cole response, unlike the bulk water \cite{Gabriel}. Such a slow, non-Debye relaxation seems to immediately imply a classical breaking of ergodicity, even if this fundamental feature does not seem to be realized in the mainstream research on the sub-Ohmic spin-boson model.

In this respect, it should be mentioned that a standard spin-boson model of electron transfer can be derived from the Hamiltonian corresponding to the Zusman model  \cite{Garg} if to assume that the reaction coordinate equilibrates very fast. However, namely, this assumption is not easy to justify for the sub-Ohmic environment. Indeed, the pertinent physical model is one of spin 1/2 (mathematically equivalent to a two-level quantum system) coupled to a harmonic oscillator (reaction coordinate), which in turn is coupled to $N$ thermal bath oscillators modeling the environment. If only the reaction coordinate relaxes much faster than the spin dynamics, it is possible to use a canonical transformation to $N+1$ harmonic oscillators at thermal equilibrium, which are directly coupled to the spin, what corresponds to the standard spin-boson model \cite{Garg}. Otherwise, this is not possible, and if the relaxation becomes asymptotically a power law, the above assumption generally becomes quite questionable indeed.

It must be stressed that the adjective ``slow'' always has a relative meaning. Perceived absolutely, it can be very misleading. For example, a typical time constant $\tau_r$ entering the Cole-Cole dielectric response expression, which corresponds to a Mittag-Leffler relaxation law in Eq. (13) below,  is for the water bound in various biological tissues in the range of $6.8-13.8$ ps \cite{Gabriel,footnote}. For fractional dynamics of various proteins, this time scale can also be in the range of $2-40$ ps, as shown both in molecular dynamics simulations and experiments \cite{Shen01,Kneller04,Calladrini08,Calladrini10,Calligari11,Calligari15}, probably due to low frequency molecular vibrations and hydrogen bond dynamics, or in the range of nanoseconds \cite{Senet08}, due to the amino acid side chain rotations. Water at the protein-solvent interface also exhibits akin anomalous relaxation and dynamics in the range of picoseconds \cite{Bizzarri02}, which is fast for common sense and everyday experience. The corresponding dielectric response should be quite stationary in the lab. The inverse of $\tau_r$ corresponds to the frequency on which the corresponding medium's degrees of freedom absorb electromagnetic energy most strongly \cite{Gabriel}. For example, myoglobin at room temperatures has a maximum in dielectric loss spectra which can be fitted by a Havriliak-Negami dielectric response with $\tau_r$ in the range of microseconds \cite{Frauenfelder09}. The corresponding medium's relaxation also follows asymptotically a power law \cite{Hilfer02}. Both Cole-Cole and Havriliak-Negami responses correspond to a power law relaxation on the time scale much larger than $\tau_r$, and this can have dramatic consequences for ET occurring in such media. The low-frequency vibrational degrees of freedom leading to such anomalous dynamics were named fractons \cite{Alexander, GranekPRL05}, while considering proteins as fractal structures of finite size at the edge of thermal stability \cite{Burioni04, Enright05, Leeuw09}, for a fixed macroconformation. However, $\tau_r$ can also belong to many orders of magnitude larger time scale, be in the range of seconds, which was also found experimentally for the slow conformational dynamics of proteins \cite{Yang03, MinPRL}. In fact, $\tau_r$ can span a huge range of variations and it typically increases exponentially with lowering temperature \cite{Frauenfelder09}. Interestingly, even coupling to high-frequency quantum vibrational modes of electron-transferring proteins can exhibit slow power-law distributed "on-off" fluctuations on the time scale from seconds to minutes \cite{BizzarriPRL05}. In this work, however, we are more interested in the relatively fast, yet anomalous dynamics with $\tau_r$ in the range from pico- to microseconds.

The generalized Zusman model of electron transfer \cite{Tang96, TangMarcusPRL} presents here a very suitable theoretical framework to address the problem of ergodic \textit{vs.} non-ergodic behavior in a semi-classical regime, with nuclei treated classically. Moreover, a very important parameter regime of non-adiabatic to solvent-controlled adiabatic transfer can be studied within the so-called contact approximation of the curve-crossing problem \cite{Zusman,Garg}. Here, the electron tunnel coupling is assumed to be much smaller than the medium's reorganization energy and smaller than the thermal energy $k_BT$ \cite{Zusman}. Nevertheless, it can still be treated non-perturbatively so that the transfer becomes independent of the strength of electronic coupling (on the ensemble level) when this coupling becomes sufficiently strong. It happens in the so-called solvent-controlled adiabatic electron transfer regime. The problem was already partially studied within a non-Markovian generalization of Zusman model by Tang and Marcus \cite{TangMarcusPRL} in the context of anomalous blinking statistics of quantum dots \cite{Fransuzov08, Pelton, Efros}, in a model of Davidson-Cole medium \cite{DavidsonCole}. However, it has never been addressed on the level of a single-trajectory description rigorously, i.e., by simulating stochastic trajectories which correspond to such a generalized Zusman model within a trajectory jump-surface analogy \cite{Tully, MuellerJCP,PCCP17}.

The significant advances of this paper are the following. First, we provide a stochastic trajectory description corresponding to the sub-Ohmic Zusman model in the contact approximation. It goes fundamentally beyond the Zusman model itself, which is formulated in the density language, on the ensemble level, and not on the level of single trajectories. Next, we revisit the Tang-Marcus theory of generalized Zusman equations in the contact approximation and confirm it in some essential detail, while deriving and representing the analytical results in a different and more insightful way. Differently from Tang and Marcus, who considered a Davidson-Cole environment \cite{DavidsonCole}, we consider a genuinely subdiffusive (generalized) Cole-Cole environment, which allows obtaining several very insightful analytical results beyond \cite{TangMarcusPRL}. For example, a novel analytical result for the population relaxation is wholly confirmed by stochastic numerics, which is a remarkable success. Our results reveal that this relaxation always has a universal power-law tail, even in the strictly non-adiabatic electron transfer regime, where the major time-course of relaxation is nearly exponential and described by the Marcus-Levich-Dogonadze (MLD) nonadiabatic rate. An analytical expression is derived for both the weight of this tail and the time point of its origin. Next, we derive also the analytical expressions for the statistics of electron transitions both on the single-trajectory and ensemble levels, which follow from the non-Markovian Zusman equations. As a great surprise, the result for the survival probability of many particles in a fixed electronic state fails against numerics beyond a strictly non-adiabatic regime of a vanishingly small electron coupling. The one for single trajectories does not fail so badly. It can agree with numerics for a substantial portion of the initial decay of survival probabilities (up to 90\%, and even more).
Moreover, it predicts the correct mean residence time, which is always finite. However, the correct tail of the distribution is very different. The theory based on generalized Zusman equations predicts two different asymptotical power-laws, one on the ensemble level and another one on the level of single trajectories. However, both tails are indeed stretched exponential, as reliable stochastic numerics reveal. By and large, a stretched exponential or Weibull distribution typifies residence time distributions within the studied model, and not a power law, by a striking contrast with the non-Markovian ensemble description. The situation here is radically different from the memoryless Ohmic case, where the non-equilibrium ensemble-based theory agrees with stochastic numerics remarkably well \cite{PCCP17}. Why non-Markovian Zusman-Tang-Marcus model formulated in the density language, on the ensemble level, profoundly fails in this respect is explained and we formulate a physically more justified ensemble approach invoking a Markovian multi-dimensional embedding of non-Markovian reaction coordinate dynamics.  

\section{Model}
We start from a standard formulation \cite{Nitzan,MayBook} of the problem of electron transfer between two diabatic, localized electronic states, $i=1,2$, with electronic energies $E_{i}(x)=\kappa (x-x_0\delta_{2,i})^2/2-\epsilon_0\delta_{2,i}$,
that depend in the Born-Oppenheimer approximation on one-dimensional nuclear reaction coordinate $x$,
which is considered as one-dimensional.
Here, for simplicity, we assume that the electronic curves are parabolic (harmonic approximation for nuclear vibrations) and have the same curvature $\kappa$ (no molecular frequency change upon electron transfer). Furthermore, $\epsilon_0$ is the difference
between electron energies at equilibrium positions of nuclei, and $x_0$ is the shift of the reaction coordinate equilibrium position upon an electronic transition. The corresponding medium's reorganization energy is $\lambda=\kappa x_0^2/2$. The diabatic electron curves cross, 
$E_1(x^*)=E_2(x^*)$, at the point, $x^*=x_0(1-\epsilon_0/\lambda)/2$, where the Born-Oppenheimer approximation is not valid (in the diabatic basis of localized states). In the vicinity of this point, electron transitions take place due to tunnel coupling 
$V_{\rm tun}$, which is assumed to be constant (Condon approximation). These are the standard assumptions, which fix a minimal and standard (thus far) model considered in this paper.
The Hamiltonian of the model formulated until this point reads $\hat H(x)=E_1(x)|1\rangle \langle 1|+E_2(x)|2\rangle \langle 2|+
V_{\rm tun}(|1\rangle \langle 2|+|2\rangle \langle 1|)$. The dynamics of the reaction coordinate $x$ will be treated classically, like the rest of the molecular degrees of freedom.  They are assumed to be at thermal equilibrium and to introduce a correlated noise and memory friction into the dynamics of the reaction coordinate, see below. Hence, we are dealing with a semi-classical description of ET, where the electron dynamics remains, however, quantum. 

Next, the probability of making a tunnel transition or a jump from one electronic curve to another can be described within the Landau-Zener-St\"uckelberger (LZS) theory \cite{Landau,Zener,Stueckelberg} as
\begin{eqnarray}
P_{\rm LZ}(v)=1-\exp\left [- f(v)\right ]\;,
\label{LZS}
\end{eqnarray}
which is a milestone result in the theory of quantum transport.
Here,
\begin{eqnarray}
f(v)=\frac{2\pi}{\hbar}\frac{|V_{\rm tun}|^2}{|(\partial \Delta E(x))/\partial x)v|_{x=x_*}},
\label{Landau}
\end{eqnarray}
with $\Delta E(x)=E_1(x)-E_2(x)=\epsilon_0-\lambda+2\lambda x/x_0$ being the difference of 
electron energies, and $v$ the reaction coordinate instant velocity at the crossing point. 
In the lowest second order approximation
in the tunnel coupling, $P_{\rm LZ}(v)\approx f(v)$. This
latter result
follows  from the Fermi's Golden Rule quantum transition rate
\begin{eqnarray}
K(x)=\frac{2\pi}{\hbar}|V_{\rm tun}|^2\delta(\Delta E(x))\;,
\label{GoldenRule}
\end{eqnarray}
applied at the crossing point. The LZS result (\ref{LZS}) is a nonperturbative result beyond the Golden Rule.

\subsection{Trajectory description}

The dynamics of the reaction coordinate will be described by a standard Kubo-Zwanzig generalized Langevin equation (GLE) \cite{KuboBook,Nitzan,ZwanzigBook}
\begin{eqnarray}
M\ddot x+\int_0^t \eta(t-t') \dot x(t')dt' +\frac{\partial E_{i}(x)}{\partial x}=\xi(t),
\label{GLE}
\end{eqnarray}
which depends on the quantum state $i$. Here, $M$ is an effective mass of the reaction coordinate, $\eta(t)$ is a memory friction kernel, and $\xi(t)$ is a correlated zero-mean Gaussian thermal noise of the environment at temperature $T$. It is completely characterized by its autocorrelation function (ACF) that is related to  the memory friction by the fluctuation-dissipation relation (FDR) named also the second fluctuation-dissipation theorem (FDT) by Kubo, $\langle \xi(t)\xi(t')\rangle =k_BT \eta(|t-t'|)$. This description comes from a multi-dimensional picture of the reaction coordinate, where an effective one-dimensional pathway, parametrized by a generalized coordinate $x$, between two stable configurations of nuclei corresponding to electronic states can be identified. Then, the rest of the molecular vibrations and possibly also the molecular degrees of freedom of a solvent surrounding the ET molecule, e.g., a protein, serve as a thermal bath for the reaction coordinate.  They introduce friction and noise, which are related by the FDT, as it follows from the main principles of equilibrium statistical mechanics, and a standard derivation of GLE dynamics from a Hamiltonian model \cite{KuboBook,ZwanzigBook,GoychukACP12}. 
In this paper, we neglect the inertial effects $M\to 0$, which corresponds to a singular model of overdamped Brownian motion with formally infinite mean-squared thermal velocity, $v_T^2=k_BT/M\to \infty$. As explained in Ref. \cite{PCCP17}, within this overdamped approximation, an effective linearization, $P_{\rm LZ}(v)\approx f(v)$, takes place, in fact, even without making explicitly a lower order expansion in $V_{\rm tun}$. In the numerical simulations though, we shall use Eq. (\ref{LZS}), to avoid an extra, explicit approximation, and to be more general on the trajectory level of description. 

Furthermore, we assume that the memory kernel consists of two parts, $\eta(t)=2\eta_0\delta(t)+\eta_\alpha t^{-\alpha}/\Gamma(1-\alpha)$, $0<\alpha<1$, where $\eta_0$ is normal Stokes friction coefficient, $\eta_\alpha$ is anomalous, fractional friction coefficient, and $\Gamma(z)$ is the well-known gamma-function. The former one corresponds the Ohmic model of the thermal bath, where the spectral bath density is linear in frequency $\omega$, $J(\omega)\propto \eta_0\omega$ \cite{WeissBook}. It corresponds to a standard exponential Debye relaxation of nuclei to equilibrium. Moreover, the latter one corresponds to a sub-Ohmic model of the thermal bath, $J(\omega)\propto \eta_\alpha\omega^\alpha$\cite{WeissBook, GoychukACP12}. For $\eta_0=0$, the relaxation dynamics of the reaction coordinate in a fixed electron state is described by the Mittag-Leffler function \cite{Mathai17}, $E_\alpha(z)=\sum_0^\infty z^n/\Gamma(1+\alpha n)$, as $\langle \delta x(t)\rangle = \delta  x(0)\theta(t)$, with relaxation function  $\theta(t)=E_\alpha[-(t/\tau_r)^\alpha]$, where $\tau_r=(\eta_\alpha/\kappa)^{1/\alpha}$ is an anomalous relaxation constant, and $\delta x(0)$ is an initial deviation from equilibrium position. $\tau_r$ will be used as a time scale in our simulations. This model corresponds \cite{GoychukPRE07Rapid} to celebrated Cole-Cole dielectric response \cite{ColeCole} commonly measured, e.g., in proteins and lipid membranes \cite{Gabriel}, where the inverse of $\tau_r$ defines a non-Debye frequency at which the medium most efficiently absorbs electromagnetic energy. With $\tau_r$ in a huge temporal range from picoseconds \cite{Shen01,Kneller04,Calladrini08,Calladrini10,Calligari11,Calligari15,GranekPRL05,Burioni04,Enright05,Leeuw09} to nanoseconds  \cite{Senet08}, and even up to seconds \cite{Yang03,MinPRL} the corresponding fractional relaxation dynamics is typical for proteins. More generally, we can have, however, a mixture of Ohmic and sub-Ohmic environments. As explained earlier \cite{GoychukPRE15}, for $\eta_0$ sufficiently small, the relaxation in a parabolic well will be almost indistinguishable from the Mittag-Leffler relaxation, and the corresponding dielectric response will be nearly Cole-Cole. We keep $\eta_0$ finite for several reasons. First, it allows justifying overdamped approximation even for $\alpha<0.4$, where it becomes questionable for $\eta_0=0$ \cite{BurovPRL08, BurovPRE08}. Second, such a normal friction component should be typically present, even when it is not dominant, e.g., for a protein in water solvent, or in the cytoplasm with dominating water content. The third reason will become clear below. 

Notice also that the model considered here differs
from one corresponding to the Davidson-Cole dielectric response that was studied by Tang and Marcus \cite{TangMarcusPRL}. The Davidson-Cole model does not yield asymptotically subdiffusion. It is not a fractional diffusion model. In fact, subdiffusion exists only on the time scale $t\ll\tau_r$ \cite{GoychukPRE07Rapid}. The asymptotic behavior in both models is very different. 
The relaxation function within the Davidson-Cole model reads $\theta(t)=\Gamma(\alpha,t/\tau_r)/\Gamma(\alpha)$ \cite{DavidsonCole,Hilfer02,GoychukPRE07Rapid}, where $\Gamma(a,z)$ is incomplete Gamma-function. It decays asymptotically exponentially, $\theta(t)\sim \exp(-t/\tau_r)/t^{1-\alpha} $, $t\gg \tau_r$, even faster than the Debye relaxation function $\theta(t)=\exp(-t/\tau_r)$.

The corresponding thermal noise $\xi(t)$ is also splitted in our model into the two parts, $\xi(t)=\xi_0(t)+\xi_\alpha(t)$, with $\langle \xi_0(t)\xi_0(t')\rangle =2k_BT\delta(t-t')$ and $\langle \xi_\alpha(t)\xi_\alpha(t')\rangle =k_BT\eta_\alpha |t-t'|^\alpha/\Gamma(1-\alpha)$. $\xi_0(t)$ is a standard white Gaussian noise (time derivative of Wiener process), whereas $\xi_\alpha(t)$ is a fractional Gaussian noise \cite{Mandelbrot68} (time derivative of fractional Brownian motion \cite{Mandelbrot68,Kolmogorov,KolmogorovTrans}). Using the notion of fractional Caputo derivative, $\frac{{\rm d}^\alpha x}{{\rm d}t^\alpha}:=\int_0^t  (t-t')^{-\alpha} \dot x(t')dt'/\Gamma(1-\alpha)$ \cite{Mathai17}, the corresponding GLE can be rewritten in the form of a fractional Langevin equation (FLE) \cite{LutzFLE,KouPRL,GoychukPRL07,GoychukPRE09,GoychukACP12,PRE13,PCCP18}
\begin{equation}
\eta_0\frac{{\rm d} x}{{\rm d}t}+\eta_\alpha\frac{{\rm d}^\alpha x}{{\rm d}t^\alpha}+\frac{\partial E_{i}(x)}{\partial x}=
 \xi_0(t)+\xi_{\alpha}(t)\;.
\label{GLE1}
\end{equation}
In numerical simulations done in the spirit of a surface hopping approach \cite{Tully,MuellerJCP,Gehlen94}, dynamics of the reaction coordinate is propagated in accordance with this equation (its finite-dimensional approximate Markovian embedding, see below) in a fixed electronic state and at each crossing of $x^*$ a jump into another electron state can occur with the above probability $P_{\rm LZ}(v)$. If an electron transition occurs, $x$ is further stochastically propagated on the another curve $E_i(x)$, until the electron jumps back, on so on, for a very long time covering huge many transitions.\\

\subsection{Generalized Zusman equations}
This trajectory model has for arbitrary $\eta(t)$ in the overdamped limit of $M\to 0$, a known ensemble counterpart. It is provided by the generalized Zusman equations \cite{Tang96,TangMarcusPRL}, considered in the contact approximation. Indeed, the
overdamped motion of the reaction coordinate in one fixed electronic state is described by the non-Markovian Fokker-Planck equation  (NMFPE), $\dot p_i(x,t)=\hat L(t) p_i(x,t)$, with a time-dependent Smoluchowski operator \cite{Hanggi77,Hanggi78,Hynes86}
\begin{eqnarray}\label{SmoluchOp}
\hat L_{i}(t)&=&D(t)\frac{\partial}{\partial x} e^{-\beta E_{i}(x)}\frac{\partial}{\partial x}
e^{\beta E_{i}(x)} \nonumber \\
&=& \frac{D(t)}{x_T^2}\frac{\partial}{\partial x}\left (x-x_0\delta_{2,i}+x_T^2
\frac{\partial}{\partial x}\right) \;\\
&:=& D(t) \hat L_i^{(0)}. \nonumber
\end{eqnarray}
Here, $\beta=1/k_BT$ is the inverse temperature, $x_T^2=k_BT/\kappa$ is the equilibrium variance of the reaction coordinate distribution in a fixed electronic state, and $D(t)$ is a time-dependent diffusion coefficient whose time-dependence expresses non-Markovian memory effects. It reads \cite{Hanggi77,Hanggi78,Hynes86}
\begin{eqnarray}\label{Dtime}
D(t)=-x_T^2\frac{d}{dt}\ln \theta(t)\;,
\end{eqnarray} 
where $\theta(t)$ is the coordinate relaxation function in a fixed electronic state,  with the Laplace-transform $\tilde \theta(s):=\int_0^{\infty}e^{-st}\theta(t)dt$ reading \cite{GoychukPRE07Rapid,GoychukACP12}
 \begin{eqnarray}\label{eta-s}
 \tilde \theta(s)=\frac{\tilde \eta(s)}{\kappa+s\tilde \eta(s)}\;,
 \end{eqnarray}
 for arbitrary memory kernel $\eta(t)$ in Eq. (\ref{GLE}) (with $M=0$). 
 For example, for the fractional dynamics in Eq. (\ref{GLE1})
with $\eta_0=0$, we have $\theta(t)=E_\alpha[-(t/\tau_r)^\alpha]$ and 
$D(t)=-x_T^2 d \ln E_\alpha[-(t/\tau_r)^\alpha]/dt$.
 It must be emphasized that such equations are known only for strictly parabolic $E_i(x)$. 
 The exact form of $D(t)$ for non-linear dynamics remains simply unknown.
From this already, one can conclude that the trajectory description given above is much more general. It can be readily generalized to a nonlinear dynamics of the reaction coordinate. The solution of NMFPE for some initial $p_i(x,t_0=0)=\delta(x-x')$ yields the well-known Green functions   \cite{Hanggi77,Hanggi78,Hynes86}
\begin{eqnarray}\label{Green}
G_i(x,t|x')=\frac{1} {\sqrt{2\pi x_T^2[1-\theta^2(t)]}}
e^{ -\frac{[x-x_0\delta_{i,2}-x'\theta(t)]^2}{2x_T^2[1-\theta^2(t)]}}\;,
\end{eqnarray}
which play an important role in the theory.

It must be mentioned that any convolution-less NMFPE with a time-dependent $D(t)$ can also be formally brought into an alternative form \cite{Hanggi77,Mukamel78},
\begin{eqnarray}\label{convolution}
\dot p_i(x,t)=\int_0^t \hat L_i^{(c)}(t-t') p_i(x,t')dt',
\end{eqnarray}
with the Laplace-transformed  $\tilde {\hat L}_{i}^{(c)}(s)$ which is related in the operator form 
as $\tilde G_i(s)=[s-\tilde {\hat L}_{i}^{(c)}(s)]^{-1}$ to
the Laplace-transformed Green-function $\tilde G_{i}(x,s|x')$ corresponding to the Smoluchowski operator $\hat L_{i}(t)$. This relation suffices for the following. We do not need to know $\hat L_i^{(c)}(t-t')$ explicitly.

 For the model in Eq. (\ref{GLE1}), the Laplace-transformed relaxation function reads \cite{GoychukPRE15}
\begin{eqnarray}\label{rfun}
\tilde \theta(s)&=&\frac{\tau_0+\tau_r (s\tau_r)^{\alpha-1}}{s\tau_0+1+(s\tau_r)^\alpha},
\end{eqnarray}
where $\tau_0=\eta_0/\kappa$, $\tau_r=(\eta_\alpha/\kappa)^{1/\alpha}$. In a particular 
case of $\alpha=0.5$, which will be studied numerically in this work, the invertion to the time domain can be easily done. It reads,
\begin{eqnarray}
\theta(t)=\frac{1}{2}\left ( 1+\frac{1}{\sqrt{1-4z}}\right )
e^{(1-\sqrt{1-4z})^2t/(4z^2\tau_r)} \nonumber \\
\times {\rm erfc}\left [ (1-\sqrt{1-4z})\sqrt{t/(4z^2\tau_r)} \right ] \nonumber \\
+\frac{1}{2}\left ( 1-\frac{1}{\sqrt{1-4z}}\right )e^{(1+\sqrt{1-4z})^2t/(4z^2\tau_r)} \nonumber \\
\times {\rm erfc}\left [ (1+\sqrt{1-4z})\sqrt{t/(4z^2\tau_r)} \right ] \; ,
\end{eqnarray} 
where $z=\tau_0/\tau_r$, and $\rm erfc$ is complementary error function.
Furthermore, for any $0<\alpha<1$,
if $\tau_0\ll \tau_r$, then relaxation follows approximately 
\begin{eqnarray}\label{theta_approx}
\theta(t)\approx E_{\alpha}[-(t/\tau_r)^\alpha], 
\end{eqnarray}
except for a small range of initial times $t< \tau_0=z\tau_r$ \cite{footnote1}.

Now we are in a position to write down generalized Zusman equations in a contact approximation 
by taking electron tunneling into account, which happens at the curve crossing point $x^*$.  For a joint probability density, $p_i(x,t)$, of electronic level populations $i$ and values $x$ of the reaction coordinate these equations 
read \cite{TangMarcusPRL},
\begin{eqnarray}
\dot p_1(x,t)&=&-K(x)[ p_1(x,t)-p_2(x,t)] +\hat L_1(t) p_{1}(x,t), \nonumber \\ 
\dot p_2(x,t)&=&K(x)[ p_1(x,t)-p_2(x,t)]+\hat L_2(t) p_{2}(x,t),
\label{Zusman}
\end{eqnarray}
where $K(x)$ is the Golden Rule expression in Eq. (\ref{GoldenRule}). It can be written as
$K(x)=v_0\delta(x-x^*)$, with  $v_0=\pi |V_{\rm tun}|^2x_0/(\hbar\lambda)$ being a tunneling velocity at the crossing point \cite{footnote2}. These are nothing else  classical anomalous diffusion-reaction equations with sink terms expressing quantum transitions from one to another electronic state. 
Based on an earlier theory of generalized Zusman equations by Tang \cite{Tang96}, these equations were introduced
by Tang and Marcus \cite{TangMarcusPRL} to study statistics of single-electron transitions in a model of  quantum dots. They appear also within a generalized Sumi-Marcus theory for a narrow reaction window approximation \cite{Sumi86,Zhu92,footnote2}. In the latter case, $x^*$ and $v_0$ have, however, a different interpretation \cite{Sumi86,Zhu92}, which we will not consider here.  Formally, equations (\ref{Zusman}) look similar to the original, memoryless Zusman equations in the contact approximation \cite{Zusman}. The difference is that the memoryless Smoluchowski operators are just replaced by ones with a time-dependent $D(t)$ that expresses non-Markovian memory effects \cite{Hanggi77,Hanggi78,Hynes86}. Beyond the contact approximation, within the four component Zusman equations, $K(x)$ should be an involved function peaked at $x^*$  whose explicit expression has been found \cite{Hartmann00} thus far for a strictly Markovian Debye model only. 
Below we solve equations (\ref{Zusman}) and compare our solution with the earlier results \cite{Tang96,TangMarcusPRL}. Moreover, our analytical solution will be tested against the numerical results of the stochastic trajectory description.  It will be shown where and why the overall approach based on a non-Markovian Fokker-Planck equation fails to describe statistics of single-electron transition events, in principle, i.e., its principal limitations will be revealed. These fundamental limitations reflect non-ergodic nature of electron transfer in markedly non-Debye environments. 
\section{Analytical theory}
\subsection{Evolution of electronic populations}
Our first goal is to derive an analytical expression for the relaxation of electronic populations $p_{1,2}(t)=\int_{-\infty}^{\infty}p_{1,2}(x,t)dx$.  We start from a formal convolution analogy of Eq. (\ref{Zusman}) written in the vector-matrix operator form and Laplace-transformed,
\begin{eqnarray}
\left [s {\mathbf I} +  {\mathbf K}(x) -\tilde {{\mathbf L}}(s)  \right ] \tilde{\mathbf P}(x,s)={\mathbf P}(x,0) .
\end{eqnarray}
Here, $\mathbf I$ is $2\times 2$ unity matrix and 
\begin{eqnarray}
\tilde{\mathbf P}(x,s)&=&\left(\begin{array}{c}
\tilde p_1(x,s)\\ \tilde p_2(x,s)\end{array}\right),
\;\;\;\tilde {\mathbf L}(s)=\left(\begin{array}{cc}\tilde {\hat L}_{1}^{(c)}(s) &0\\0& \tilde {\hat L}_{2}^{(c)}(s)
\end{array}\right),\nonumber \\
{\mathbf K}(x)&=&K(x)\left(\begin{array}{cc}1&-1\\-1&
1\end{array}\right),  \mathbf P(x,0)=\left(\begin{array}{c}
 p_1(x,0)\\ p_2(x,0)\end{array}\right)\;.
\end{eqnarray}
All the corresponding  Laplace-transforms are denoted as the original quantities with tilde and Laplace variable $s$ instead of time variable $t$. Next, we proceed closely to Ref. \cite{Hartmann00} and use a projection operator $\mathbf \Pi$ whose action on arbitrary function $f(x)$ is defined by
\begin{eqnarray}\label{projector}
{\mathbf \Pi}f(x) & = &
\left(\begin{array}{cc}p_1^{(eq)}(x) &0\\0&p_2^{(eq)}(x)\end{array} \right)
\int_{-\infty}^{\infty}f(x)dx \nonumber \\
& = & {\mathbf P}_{\rm eq}(x)\int_{-\infty}^{\infty}f(x)dx, \nonumber \\
\end{eqnarray}
where $p_i^{(eq)}(x)=\exp[-(x-x_0\delta_{i,2})^2/(2x_T^2)]/\sqrt{2\pi x_T^2}$ are the equilibrium distributions of the reaction coordinate in the fixed electronic states. It is easy to check that $\mathbf \Pi^2=\mathbf \Pi$, and $\mathbf \Pi \tilde{\mathbf P}(x,s)= {\mathbf P}_{\rm eq}(x) \tilde{\mathbf p}(s)$, where $\tilde{\mathbf p}(s)=[\tilde p_1(s), \tilde p_2(s)]^T$ is vector of Laplace-transformed electronic populations. This allows to split 
$\tilde{\mathbf P}(x,s)$ as $\tilde{\mathbf P}(x,s)= {\mathbf P}_{\rm eq}(x) \tilde{\mathbf p}(s)+\tilde{\mathbf P}_1(x,s)$, where $\tilde{\mathbf P}_1(x,s)={\mathbf Q} \tilde{\mathbf P}(x,s)$ is an orthogonal vector and ${\mathbf Q}=\mathbf I-\mathbf \Pi$ is a complementary projector, $\mathbf \Pi \mathbf Q=\mathbf Q \mathbf \Pi=0$. Using standard operations with projection operators and such properties as $\tilde {{\mathbf L}}(s) {\mathbf P}_{\rm eq}(x)=0$ allows to exclude the irrelevant part $\tilde{\mathbf P}_1(x,s)$.  After some standard algebra,
we obtain  the following exact result
\begin{equation}\label{laplace2}
[{\mathbf  k}(s)+s {\mathbf I}]{\mathbf p}(s)={\bf p}(0),
\end{equation}
with the matrix
\begin{equation}\label{k-s}
{\mathbf k}(s)={\mathbf P}_{\rm eq}^{-1}{\mathbf \Pi}{\mathbf K}\left ({\mathbf I}-
[s{\mathbf I}+{\mathbf Q}({\mathbf K}- \tilde {{\mathbf L}}(s))]^{-1}
{\mathbf Q}{\mathbf K}\right)  {\mathbf P}_{\rm eq}.
\end{equation}
This result holds for the class of initial preparations with equilibrated reaction coordinate, $p_i(x,0)=p_i^{(eq)}(x)p_i(0),\; p_1(0)+p_2(0)=1$. Next, using $\mathbf Q \tilde {{\mathbf L}}(s)=\tilde {{\mathbf L}}(s)$ and formal operator expansions like 
$[\hat A+\hat B]^{-1}=[\hat A(1+\hat A^{-1}\hat B)]^{-1}=[1+\hat A^{-1}\hat B]^{-1}\hat A^{-1}=\sum_{n=0}^\infty (-1)^n(\hat A^{-1}\hat B)^n\hat A^{-1}$, with $\hat A= s{\mathbf I}-\tilde {{\mathbf L}}(s)$, and $\hat B={\mathbf Q}{\mathbf K}$,
the above result can formally exactly be represented as
\begin{equation}\label{k-s2}
{\mathbf k}(s)={\mathbf P}_{\rm eq}^{-1}{\mathbf \Pi}{\mathbf K}\left [{\mathbf I}-
\sum_n^{\infty}(-1)^n [\tilde { {\mathbf G}}(s){\mathbf Q}{\mathbf K}]^{n+1}\right ]  {\mathbf P}_{\rm eq},
\end{equation}
where $\tilde {{\mathbf G}}(s)=[s{\mathbf I}-\tilde {{\mathbf L}}(s)]^{-1}$ is the Laplace-transformed Green function operator. In the coordinate representation, its components read $\tilde G_{ij}(x,s|x')=\delta_{ij}\tilde G_i(x,s|x')$, with $\tilde G_i(x,s|x')$ being the Laplace-transformed Green-function in Eq. (\ref{Green}), which is well-known. In this respect, action of operator 
$\tilde{ G}_i(s)$ on any function $f(x)$ is defined by the integral $\int_{-\infty}^{\infty}
\tilde G_i(x,s|x')f(x')dx'$. Within the considered model with $K(x)=v_0\delta(x-x^*)$, the multiple integrals entering Eq. (\ref{k-s2}) can be reduced to powers of one-dimensional integrals and the resulting geometric series can be summed up exactly. We obtain the exact result,
\begin{equation}\label{rate-matrix}
{\mathbf k}(s)=\Big[{\mathbf I}+{\mathbf K}^{(na)}{\tilde{\mathbf T}}(s) \Big]^{-1}{\mathbf K}^{(na)}.
\end{equation}
Here, the elements of the matrices
\begin{eqnarray}
{\mathbf K}^{(na)}& =&\left(\begin{array}{cc}k_1^{(na)}&-k_2^{(na)}\\-k_1^{(na)}&k_2^{(na)}\end{array}\right),\nonumber \\
\;\tilde{{\mathbf T}}(s)&=&\left(\begin{array}{cc}\tilde \tau_1(s)&0\\0&\tilde \tau_2(s)\end{array}\right),
\end{eqnarray}
are defined by the integral relations
\begin{eqnarray}\label{na-rate}
k_{1,2}^{(na)}&=&\int_{-\infty}^{+\infty}K_{}(x)p_{1,2}^{(eq)}(x)dx\nonumber \\
            &=&\frac{2\pi V_{\rm tun}^2 }{\hbar\sqrt{4\pi\lambda k_BT}}
  e^{- \frac{E^{(a)}_{1,2}}{k_BT}}\;,
\end{eqnarray}
and
\begin{eqnarray}\label{tau(s)}
\tilde \tau_{1,2}(s)&=&\int_{0}^{\infty}dt\;e^{-st}
\left[G_{1,2}(x^*,t|x^*)/p_{1,2}^{(eq)}(x^*)-1\right]\nonumber\\
&=& \tilde G_{1,2}(x^*,s|x^*)/p_{1,2}^{(eq)}(x^*)-1/s\;.
\end{eqnarray}
Eq. (\ref{na-rate}) is the celebrated Marcus-Levich-Dogonadze expression \cite{Marcus60,Levich59,Hush96} for the rate of non-adiabatic electron transfer. Here, $E^{(a)}_{1,2}=(\epsilon_0\mp\lambda)^2/(4\lambda)$ are the activation energies displaying a parabolic dependence on the energy bias $\epsilon_0$ (the famous Marcus parabola). The very fact that this is a quantum rate, despite it is often named classical, is expressed by the quantum tunneling prefactor in Eq. (\ref{na-rate}). Furthermore, the limit $\lim_{s\to 0}\tilde \tau_i(s)=\tau_i^{ad}=1/k_i^{ad}$, in Eq. (\ref{tau(s)}), when exists, yields the inverse of adiabatic Marcus-Hush \cite{Marcus56,Hush58,Hush96} rates of electron transfer $k_i^{ad}$ (for Debye solvents), or their generalizations (beyond Debye solvents). As a result, the dynamics of electronic populations is governed by the generalized master equations (GMEs) reading
\begin{eqnarray}\label{main1}
\dot p_1(t)& = &-\int_0^{t}k_1(t-t')p_1(t')dt'\\
& + &\int_0^{t}k_2(t-t')p_2(t')dt',\nonumber \\
\dot p_2(t)& = &-\dot p_1(t) \nonumber
\end{eqnarray}
with the memory kernels defined by their Laplace-transforms
\begin{eqnarray}\label{analyt}
 \tilde k_{ i}(s) = \frac{k_{ i}^{(na)}}{1+\tilde \tau_1(s)k_{ 1}^{(na)}+
\tilde \tau_2(s)k_{ 2}^{(na)} }\;.
 \end{eqnarray}
This is the first profound result of this work. When exist, the (generalized) Zusman rates of electron transfer read $k_{1,2}=\tilde k_{1,2}(0)$. For example, in the case of Davidson-Cole solvents such rates do exist and a Markovian approximation of the relaxation dynamics can be done on the time scale $t\gg \tau_r$, for sufficiently high activation barriers $E^{(a)}_{1,2}\gg k_BT$.
In our case of subdiffusive reaction coordinate, however, $\lim_{s\to 0} \tilde \tau_i(s)=\infty$, see below, and this has dramatic consequences for ET transfer kinetics is such subdiffusive environments because a Markovian approximation to (\ref{main1}) is generally simply wrong. Namely, in the Appendix A, it is shown that the asymptotic behavior of 
$\tilde \tau_{1,2}(s)$ for $0<\alpha<1$ in the limit $s\to 0$ and for sufficiently large activation barriers (over several $k_BT$) is 
\begin{eqnarray}\label{approx}
\tilde \tau_{1,2}(s)\sim 2r_{1,2}\tau_r(s\tau_r)^{\alpha-1},
\end{eqnarray}
where $r_{1,2}=E^{(a)}_{1,2}/(k_BT)$ is activation energy in the units of $k_BT$. 
On the other hand, the asymptotic behavior of $\tilde \tau_{1,2}(s)$ for large $s\tau_r\gg 1$ is
\begin{eqnarray}\label{approx2}
\tilde \tau_{1,2}(s)\sim \sqrt{\pi\frac{\tilde \eta_0}{2}}\left (e^{r_{1,2}}-1\right)\tau_r(s\tau_r)^{-1/2},
\end{eqnarray}
universally, for any $\alpha$, whenever $\eta_0\neq 0$. This asymptotics is very important for the statistics of single electron transitions.
Here, $\tilde \eta_0=\eta_0/(\eta_\alpha\tau_r^{1-\alpha})=z$ is a 
scaled normal friction coefficient.
  By a comparison of (\ref{approx}) and (\ref{approx2}) one can see that the both asymptotics coincide only for $\alpha=0.5$ and for $\sqrt{\frac{\tilde \eta_0}{2}}\left (e^{r_{1,2}}-1\right) =2r_{1,2}/\sqrt{\pi}$. Hence, only for a symmetric case, 
  $r_1=r_2=r$ and $\alpha=0.5$ one can choose $r$ and $\tilde \eta_0$ so that (\ref{approx}) or (\ref{approx2}) can work approximately uniformly for any $s$. Fig. \ref{FigA1}, a, serves to illustrate such a case which offers a possibility for nice analytical expressions, see below.
This is actually the third reason for choosing the model with a finite $\eta_0$. With 
$\eta_0=0$, the large-$s$ asymptotics of $\tilde \tau_{1,2}(s)$ is
\begin{eqnarray}\label{approx3}
\tilde \tau_{1,2}(s)\sim \frac{\Gamma(1-\alpha/2)\sqrt{\Gamma(1+\alpha)/2}\left (e^{r_{1,2}}-1\right)\tau_r}{(s\tau_r)^{1-\alpha/2}}\;.
\end{eqnarray}
In this case, the short and long time asymptotics can never coincide. It will be studied in detail elsewhere.

Furthermore, it is easy to show that the relaxation of populations follows
\begin{eqnarray}
p_{1,2}(t)=p_{1,2}^{(eq)}+[p_{1,2}(0)-p_{1,2}^{(eq)}]R(t),
\end{eqnarray}
where $p_{2,1}^{(eq)}=k_{ 1,2}^{(na)}/[k_{ 1}^{(na)}+k_{ 2}^{(na)}]=1/[1+\exp(\pm \epsilon_0/(k_BT))]$ are equilibrium populations and $R(t)$ is a population relaxation function with the Laplace-transform reading
\begin{eqnarray}\label{analytic1}
 \tilde R(s) &=& \frac{1}{s+\tilde k_{ 1}(s)+\tilde k_{ 2}(s)}\nonumber \\ &=&
 \frac{1}{s+\frac{k_{ 1}^{(na)}+k_{ 2}^{(na)}}{1+\tilde \tau_1(s)k_{ 1}^{(na)}+
\tilde\tau_2(s)k_{ 2}^{(na)} } }\;.
 \end{eqnarray}
This general result looks formally equivalent to one obtained first by Tang in a very different way \cite{Tang96}. 
However, Tang used a very different $\tilde G_{1,2}(x^*,s|x^*)$ \cite{Tang96}, and the results are equivalent in fact only for Debye solvents. 
Importance of this general result reaches beyond the particular GLE model of this work. For example, with the Green function of the fractional Fokker-Planck equation \cite{MetzlerPRL99}, it corresponds to a matching non-Markovian generalization of Zusman model. Moreover, other generalized Fokker-Planck equation descriptions based on continuous time random walks with finite mean residence times in traps
\cite{LeeJCP03,GoychukPRE12} can be used. These are, however, the models beyond the scope of this work.

 The Green-function in this work is related to one used by Tang and Marcus \cite{TangMarcusPRL}, who did not study, however, the relaxation of electronic populations. Moreover, they considered a Davidson-Cole medium, while we consider a Cole-Cole medium, which is another profound point of difference.  
Let us consider limiting cases of this expression for the model under study. 

\subsubsection{The limit of solvent-controlled adiabatic transfer}

First, we consider the formal limit of $V_{\rm tun}\to \infty$, in which we obtain $\tilde k_{1}(s)\approx k_1^{(ad)}(s)=
 1/[\tilde \tau_1(s)+\tilde \tau_2(s)\exp(-\epsilon_0/(k_BT))]$, $\tilde k_{2}(s)\approx k_2^{(ad)}(s)=
k_1^{(ad)}(s) \exp[-\epsilon_0/k_BT)]$, and 
\begin{eqnarray}\label{analytic1a}
 \tilde R(s) &=& \frac{\tilde \tau_1(s)p_2^{(eq)}+\tilde \tau_2(s)p_1^{(eq)}}{1+s[\tilde \tau_1(s)p_2^{(eq)}+\tilde \tau_2(s)p_1^{(eq)}] }\;,
 \end{eqnarray}
 It must be mentioned once again that physically $V_{\rm tun}$ has to be much smaller than $\lambda$  and do not exceed $k_BT$ in this solved-controlled adiabatic regime. Otherwise, the considered model cannot be physically justified. The result in Eq. (\ref{analytic1a}) is the second important result of this paper. It also solves the problem of overdamped classical anomalous relaxation with arbitrary kernel $\eta(t)$ in a cusp-like bistable potential consisting of two pieces of parabolas of equal curvature. It is so because in the limit $v_0\to\infty$ the particle crosses the boundary between two domains of attraction with the probability one, once it arrives at the boundary (absorbing boundary).  It must be, however, also emphasized that this result does not describe a typical time scale of transitions of \textit{single} particles between two domains of attraction (which does exist!) because of a broken (!) ergodicity, see below: the ensemble and single-trajectory descriptions are fundamentally different. It describes how the particles redistribute between two attraction domains, all starting, e.g. in one of them, whereas approaching an equilibrium distribution (equipartition in the symmetric case). Each particle crosses the boundary huge many times during this equilibration process.  An especially insightful and beautiful result is obtained when both asymptotics, (\ref{approx}) and (\ref{approx2}), coincide. 
Then, GME (\ref{main1}) can approximately be written as a fractional master equation \cite{Hilfer95,Metzler00}
 \begin{eqnarray}\label{main1a}
\dot p_1(t)& = &- \sideset{_{0}}{_t}{\mathop{\hat D}^{1-\alpha}} \left [ k_{\alpha,1}p_1(t)
 -k_{\alpha,2}p_2(t)\right ],  \\
p_2(t)& = &1- p_1(t), \nonumber
\end{eqnarray}
with fractional rates $k_{\alpha,1}=1/\left [2\tau_r (r_1+r_2e^{-\epsilon_0/k_BT})  \right ]^\alpha$ 
$k_{\alpha,2}=k_{\alpha,1}\exp[-\alpha\epsilon_0/k_BT)]$ and fractional Riemann-Liouville time derivative \cite{Mathai17}
\begin{eqnarray}\label{RL} 
\sideset{_{0}}{_t}{\mathop{\hat D}^{1-\alpha}}p(t):=\frac{1}{\Gamma(\alpha)}
\frac{d}{dt }\int_{0}^t dt' \frac{p(t')}{(t-t')^{1-\alpha}}\;.
\end{eqnarray}
 This remarkable form, which, anyway, is valid, in fact, only for $\alpha=0.5$, the symmetric case, $\epsilon_0=0$
 and a special choice of the pair $\eta_0,r$ (see above),
  can, however, be also very misleading. One has to be very careful with it because a perplexed reader might attribute Eq. (\ref{main1a}) to a non-stationary continuous time random walk \cite{Hughes,Scher75,Shlesinger74,Metzler00,MetzlerPRL99,GoychukRapid06} with divergent mean residence times in the traps of a rough potential landscape for the reaction coordinate. It corresponds, however, quite on the contrary, to the \textit{stationary}, equilibrated dynamics of the reaction coordinate.
Namely, such surface analogies lead to two very different ``fractional'' dynamics in the literature, which might look perplexingly very similar \cite{GoychukFractDyn11}. In this case, we have
\begin{eqnarray}\label{analytic1b}
 \tilde R(s) &=& \frac{\tau_{ad}(s\tau_{ad})^{\alpha-1}}{1+(s\tau_{ad})^{\alpha} }\;,
 \end{eqnarray}
 where we introduced a scaling relaxation constant
 \begin{eqnarray}
 \tau_{ad}=\tau_r\left (2r_1p_2^{(eq)}+2r_2p_1^{(eq)}\right )^{1/\alpha}\;.
 \end{eqnarray}
 The result in Eq. (\ref{analytic1a}) inverted to the time-domain reads $R(t)=E_{\alpha}[-(t/\tau_{ad})^\alpha]$, i.e. it is described by the Mittag-Leffler relaxation function, precisely so as the relaxation of the reaction coordinate in the considered Cole-Cole solvent ($\tau_0\ll\tau_r$), but with a very different scaling time $\tau_{ad}$. The striking feature is that $\tau_{ad}$ scales not exponentially with the height of the activation barrier and temperature, but as a power law. For symmetric case, 
 \begin{eqnarray}\label{newscale}
 \tau_{ad}=\tau_r\left (2E^{(a)}/k_BT \right)^{1/\alpha}.
 \end{eqnarray}
 This result is significant. In the case of rate processes, such power-law dependencies are usually attributed to quantum mechanical effects \cite{WeissBook}. In the present case, however, it has nothing to do with quantum mechanics.

\subsubsection{Nonadiabatic ET}

Next, it worth to notice that the inversion of $\tilde R(s)$ to the time domain can be done precisely for $\alpha=0.5$, for any $V_{\rm tun}$, within the approximation (\ref{approx})  taken for granted uniformly. Namely, such a case will be treated numerically below. This inversion reads,
\begin{eqnarray}\label{Rexact}
R(t)&=&\frac{1}{2}\left ( 1+\frac{\kappa_{ad}}{\sqrt{\kappa_{ad}^2-4\kappa_{ad}}}\right ) E_{1/2}(-\zeta_1\sqrt{t/\tau_{ad}}) \\
&+&\frac{1}{2}\left ( 1-\frac{\kappa_{ad}}{\sqrt{\kappa_{ad}^2-4\kappa_{ad}}}\right ) E_{1/2}(-\zeta_2\sqrt{t/\tau_{ad}}) \nonumber
\end{eqnarray}
where $E_{1/2}(-z)=e^{z^2}{\rm erfc}(z)$ is Mittag-Leffler function of index $1/2$ expressed via the complimentary error function. Furthermore, $\kappa_{ad}=k_{na}\tau_{ad}$ is an adiabaticity parameter, and $\zeta_{1,2}=\kappa_{ad}\left (1\mp\sqrt{1-4/\kappa_{ad}} \right)/2$. Here, $k_{na}=k_1^{(na)}+k_2^{(na)}$ is the total nonadibatic rate.
This is an important result to be checked against numerics because of its simplicity and the insights it provides.  In the adiabatic transfer regime, $\kappa_{ad}\gg 1$, $\zeta_1\approx 1$ and $\zeta_2\approx \kappa_{ad}$.
In this case, $R(t)\approx E_{1/2}(-\sqrt{t/\tau_{ad}})$. For $\kappa_{ad}<4$, $\zeta_1$ and $\zeta_2$ are complex-conjugated with the real part $1/2$.

Furthermore, the asymptotic behavior of $R(t)$ is universal, as follows from (\ref{approx}):
\begin{eqnarray}\label{puzzle}
R(t)\sim \frac{1}{\Gamma(1-\alpha)}\left (\frac{\tau_{ad}}{t} \right)^\alpha\;, t\to\infty 
\end{eqnarray}
This behavior in Eq. (\ref{puzzle}) has not been found earlier for ET in non-Debye environments. The algebraic scaling of the tail with time, $R\propto 1/t^\alpha$, apart from a nontrivial time constant $\tau_{ad}$ entering it, reflects the behavior of the autocorrelation function of fractional Gaussian noise $\xi_\alpha(t)$. It worth noting that a similar heavy tail was also found in the relaxation of a two-level quantum-mechanical system driven by a very different two-state stationary non-Markovian noise whose autocorrelation function exhibits, however, the same power-law scaling in its asymptotic decay \cite{GoychukCP06}. Hence, it seems to be a generic feature, independently of the noise amplitude statistics, being primarily determined by the scaling of its ACF. Given this result and that $\lim_{s\to 0} k_{1,2}^{(na)}\tilde \tau_{1,2}(s)=\infty$, it seems first very questionable that non-adiabatic ET transfer regime with 
\begin{eqnarray}\label{nad}
R(t)=\exp[-k_{na}t],
\end{eqnarray}
can exist at all in a Cole-Cole medium. Indeed, the initial behavior of $R(t)$ in adiabatic transfer regime is stretched exponential, $R(t)\approx \exp[-(t/t_{in})^\gamma]$, with $\gamma=\alpha$ and $t_{in}=\tau_{ad} \Gamma(1+\alpha)^{1/\alpha}$. However, with decreasing tunnel coupling $V_{\rm tun}$, the stretching power-law exponent $\gamma$ gradually approaches unity. Hence, with ever smaller $V_{\rm tun}$, the initial regime (\ref{nad}), where ET has a nonadiabatic character, is not only gradually established, but it can cover over 90\% of the population transfer. Nevertheless, the residual power-law tail starts at some transition time $t_c$, which can be found from an approximate matching condition
\begin{eqnarray}\label{cond}
\exp[-k_{na}t_c]=\frac{1}{\Gamma(1-\alpha)}\left (\frac{\tau_{ad}}{t_c} \right)^\alpha,
\end{eqnarray}
solved for a large $t_c/\tau_{ad}$. 
The corresponding real solution reads
\begin{eqnarray}\label{Lambert}
t_c=-\tau_{ad}\frac{\alpha}{\kappa_{ad}}{\rm LambertW}\left(-1,-\frac{\kappa_{ad}}{\alpha\Gamma(1-\alpha)^{1/\alpha}}
  \right),
  \end{eqnarray}
 where ${\rm LambertW}(-1,x)$ is a $-1$ branch of the Lambert special function \cite{Corless}. In the nonadiabatic anomalous ET regime, the adiabaticity parameter $\kappa_{ad}\ll 1$. For example, for $\alpha=0.5$ and $\kappa_{ad}=0.01$, $t_c\approx 350.155\;\tau_{ad}$, and the corresponding $R(t_c)\approx 0.0301$, i.e. about 97\% of population relaxation is nearly exponential and well described by the non-adiabatic MLD rate. However, the rest 3\% follows an algebraically slow approach to equilibrium. Another example, for $\kappa_{ad}=0.001$,  $t_c\approx 4811.776 \;\tau_{ad}$, and $R(t_c)\approx 0.0081$.  Clearly, in a deeply nonadiabatic ET regime, a heavy tail with such a small initial amplitude can be masked by the population fluctuations, see below, be buried in them, and, hence, not detectable. Therefore, beyond any doubts, a non-adiabatic ET does exist in a sense described even in dynamically anomalously slow environments.

\subsection{Survival probabilities in electronic states: an equilibrium ensemble perspective}

Let us now pose the question: What is the survival probability $F_i(t)$ of electron in the state $i$ before it switches to another state for the \textit{first} time?  To answer this question one should forbid 
the return of electron after it made the transition, i.e. to put to zero either $k_1(t)\to 0$ or 
$k_2(t)\to 0$ in Eq. (\ref{main1}), and either $k_1^{(na)}\to 0$, or $k_2^{(na)}\to 0$ in the denominator of Eq. (\ref{analyt}). Then, the answer follows immediately in the Laplace-space from Eq. (\ref{analyt}) :
\begin{eqnarray}\label{analytic2}
 \tilde F_{i}^{(ens)}(s) =
 \frac{1+\tilde \tau_i(s)k_{ i}^{(na)} }{s [1+\tilde \tau_i(s)k_{ i}^{(na)} ]+k_{ i}^{(na)}}\;.
 \end{eqnarray}
For the model under study, the long-time behavior of $F_{i}^{(ens)}(t)$ displays the same universal feature,
\begin{eqnarray}\label{asymp1}
F_i^{(ens)}(t)\sim \frac{1}{\Gamma(1-\alpha)}\left (\frac{\tau_{i,ad}}{t} \right)^\alpha\;, t\to\infty 
\end{eqnarray}
but with a different constant $\tau_{i, ad}=\tau_r\left (E_i^{(a)}/k_BT \right)^{1/\alpha}$. 
Likewise, all the above discussed features of $R(t)$ apply to $F_{1,2}^{(ens)}(t)$ upon putting $\tau_{2, ad}\to 0$,
or $\tau_{1, ad}\to 0$, respectively, in the corresponding expressions for $R(t)$.  The most striking feature of the corresponding RTD, $\psi_i(t)=-dF_i(t)/dt\propto 1/t^{1+\alpha}$ is that it does not have a finite mean. For the case of $\alpha=0.5$ and for the parameters where the approximation (\ref{approx}) works uniformly, the corresponding $F_i^{(ens)}(t)$ are given by the rhs of Eq. (\ref{Rexact}), with $\tau_{i,ad}$ instead of $\tau_{ad}$, and $\kappa_{i,ad}$  replaced by $\kappa_{i,ad}=k_i^{(na)}\tau_{i,ad}$. Accordingly, $\zeta_{1,2}^{(i)}=\kappa_{i,ad}\left (1\mp\sqrt{1-4/\kappa_{i,ad}} \right)/2$, instead of $\zeta_{1,2}$.

\subsection{Statistics of single-electron transitions}

The result in Eq. (\ref{analytic2}) does not describe, however, the statistics of single-electron transitions, $F_i^{(sgl)}(t)$. Indeed, each jump of an electron occurs at the same (in the contact approximation)  very non-equilibrium value $x^*$ of the reaction coordinate. This feature is very different from the assumption about a thermally equilibrated reaction coordinate resulting in Eq. (\ref{analytic2}).  In fact, the quantum nature of electron transitions is indispensable even in the classical adiabatic ET regime, when it is considered on the level of single trajectories. This fact leads to a quantum breaking of ergodicity even in Debye solvents. To derive the statistics of single-electron transitions from generalized Zusman equations one must think differently \cite{TangJCP, TangMarcusPRL, PCCP17}. Indeed, let  an electron to start in the state $i$ at time $t_0=0$ with the reaction coordinate fixed at $x'$. Then, at time $t$,
$p_i(x,t)$ obeys an integral equation \cite{WilemskiFixman}
\begin{eqnarray}\label{basic1}
&&p_i(x,t)=G_i(x,t|x')\nonumber \\
&- &\int_0^tdt'\int_{-\infty}^{\infty}dx'G_i(x,t-t'|x')K(x')p_i(x',t').
\end{eqnarray}
We are interested in the survival probability in this state and therefore consider only sink $K(x)$ out of this state. With $K(x)=v_0\delta(x-x^*)$ this yields
\begin{eqnarray}\label{basic2}
p_i(x,t)& = &G_i(x,t|x')\nonumber \\
&-&v_0\int_0^tdt'G_i(x,t-t'|x^*)p(x^*,t').
\end{eqnarray}
The Laplace-transform of this equation gives
\begin{eqnarray}\label{basic3}
\tilde p_i(x,s) = \tilde G_i(x,s|x')-v_0\tilde G_i(x,s|x^*)\tilde p(x^*,s), 
\end{eqnarray}
and from it one can find $\tilde p_i(x,s)$,
\begin{eqnarray}\label{basic4}
\tilde p(x^*,s)=\frac{\tilde G_i(x^*,s|x')}{1+v_0 \tilde G_i(x^*,s|x^*)}.
\end{eqnarray}
The survival probability in the state $i$ is $F_i(t)=\int_{-\infty}^{\infty}p_i(x,t)dx$. Hence, from Eqs. 
(\ref{basic3}), (\ref{basic4}), and using normalization of Green function, $\int G_i(x,t|x')dx=1$, we find
\begin{eqnarray}\label{basic5}
\tilde F_i(s)&=&\frac{1}{s}\Big [1-v_0 \tilde p_i(x^*,s)\Big ]\nonumber \\
&=&\frac{1+v_0 [\tilde G_i(x^*,s|x^*)-\tilde G_i(x^*,s|x')]}{s+v_0 s\tilde G_i(x^*,s|x^*)}.
\end{eqnarray}
Now, if the initial $x'$ is taken from the equilibrium distribution of the reaction coordinate, then one must replace $G(x,t|x')$ with $p_i^{(eq)}(x)$ in the first line of Eqs. (\ref{basic1}), (\ref{basic2}) and also $\tilde G_i(x^*,s|x')$ with $p_i^{(eq)}(x^*)/s$ in Eq. (\ref{basic5}). Then, upon taking Eq. (\ref{tau(s)}) into account and the fact that $k_i^{(na)}=\nu_0p_i^{(eq)}(x^*)$ we immediately reproduce the result in Eq. (\ref{analytic2}). This is just another way to derive it. However, for the statistics of single trajectories one must take $G_i(x^*,s|x')=G_i(x^*,s|x^*)$ in Eq. (\ref{basic5}), which yields
\begin{eqnarray}\label{analytic3}
 \tilde F_{i}^{(sgl)}(s) =
 \frac{1}{s [1+\tilde \tau_i(s)k_{ i}^{(na)} ]+k_{ i}^{(na)}}\;
 \end{eqnarray}
instead of (\ref{analytic2}). The difference is, in fact, huge. First of all, with Eq. (\ref{approx}) in (\ref{analytic3}), one can see immediately that the mean residence time in the electronic states 
$\langle \tau_i \rangle=\lim_{s\to 0}\tilde F_{i}^{(sgl)}(s)$, not only exists, but it equals always the inverse MLD rate, $\langle \tau_i \rangle=1/k_i^{(na)}$. This is a striking result. It shows how misleading can an equilibrium ensemble theory perspective be for the single-trajectory statistics! The result in Eq. (\ref{analytic3}) is equivalent to one in Eq. (2)
of Ref. \cite{TangMarcusPRL} by Tang and Marcus for the RTD $\tilde \psi_i(s)$ therein (our notations are different), which can be obtained as $\tilde \psi_i(s)=1-s \tilde F_{i}^{(sgl)}(s)=\tilde p_i(x^*,s)$.
However, our form is better because it allows avoiding some pitfalls in the analysis possible especially in the case of finite adiabatic times $\tilde \tau_{1,2}(0)$.
It predicts a very different from the equilibrium ensemble perspective power-law for the electron RTDs, 
$\psi_i(t)\propto 1/t^{2+\alpha}$, for large sojourn time intervals.
Indeed, with (\ref{approx}) in (\ref{analytic3}) one can show upon using some identical transformations and a Tauberian theorem \cite{Doetsch} that 
\begin{eqnarray}\label{asymp2}
F_i^{(sgl)}(t)\sim\frac{1}{k_i^{(na)}\tau_r} \frac{2\alpha E_{i}^{(a)}}{\Gamma(1-\alpha)k_BT}\left (\frac{\tau_r}{t} \right)^{1+\alpha}\;, 
\end{eqnarray}
for $t\gg \tau_r$, and $\psi_i(t)$ is a negative derivative of this result. Notice that
(\ref{asymp2}) is very different from (\ref{asymp1}). 

However, we will show soon that this prediction is wrong: For the considered non-Markovian dynamics, the tail of the distribution is very different. It is a stretched exponential, and the generalized Zusman equations fail to describe it. The situation here is very different from the Markovian dynamics, where Eq. (\ref{analytic3}) was very successful in predicting the statistics of single trajectories \cite{PCCP17}. In the present case, the quantum breaking of ergodicity combines with the classical one, caused by an algebraically slow dynamics of the reaction coordinate, which leads to a new dimension of complexity.

In an important particular case of $\alpha=0.5$ and for (\ref{approx}) used for all $s$, 
\begin{eqnarray}\label{Fexact}
F_i^{(sgl)}(t)&=&\frac{1}{2}\left ( 1+\frac{\kappa_{i,ad}}{\sqrt{\kappa_{1,ad}^2-4\kappa_{i,ad}}}\right )\nonumber \\ &\times&E_{1/2}\left (-\zeta_2^{(i)}\sqrt{t/\tau_{i,ad}}\right ) \\
&+&\frac{1}{2}\left ( 1-\frac{\kappa_{i,ad}}{\sqrt{\kappa_{i,ad}^2-4\kappa_{i,ad}}}\right )\nonumber \\ &\times& E_{1/2}\left (-\zeta_1^{(i)}\sqrt{t/\tau_{i,ad}}\right ) \nonumber\;.
\end{eqnarray}
The formal difference with the corresponding expression for $F_i^{(ens)}(t)$ seems really small and subtle. However, the consequences are really profound! Indeed, in the adiabatic transfer regime  the main behavior of $F_i^{(sgl)}(t)$ covering about 70-90\% of survival probability initially is given by 
$F_i^{(sgl)}(t)\approx E_{1/2}\left (-\sqrt{t/\tau_{i,sgl}}\right )$, where 
\begin{eqnarray}\label{tau_single}
\tau_{i,sgl}&\approx& \tau_{i,ad}/\kappa_{i,ad}^2=1/[(k_i^{(na)})^2\tau_{i,ad}]\;.
\end{eqnarray} 
For example,
for $\kappa_{i,ad}=10$, $\tau_{i,sgl}$ is 100 times (!) smaller than $\tau_{i,ad}$ entering formally the same approximate (for the initial times) expression for $F_i^{(ens)}(t)$ with the only difference: $\tau_{i,ad}$
instead of $\tau_{i,sgl}$.
Furthermore, to be more general and to go beyond a very restrictive case of coinciding (\ref{approx}) and (\ref{approx2}), we should use a different from 
$\tau_{i,ad}=4\tau_r r_{1,2}^2$ expression for $\tau_{i,ad}$. Namely, one should use the one stemming from the short-time/large$-s$ asymptotics in Eq. (\ref{approx2}) that yields 
$\tau_{i,ad}'\approx (\pi/2)\tau_r\tilde \eta_0 e^{2r_{1,2}}$, for $r_{1,2}\gg 1$.
With this in Eq. (\ref{tau_single}) we obtain 
\begin{eqnarray}\label{tau_single2}
\tau_{i,sgl}\approx \frac{2\hbar^2 \lambda k_BT }{ \pi^2V_{\rm tun}^4\tau_0 }.
\end{eqnarray}
Notice, that the Debye relaxation time $\tau_0$ enters this expression, and not $\tau_r$. Eq. (\ref{tau_single2}) coincides with one by Tang and Marcus in Ref. \cite{TangJCP}. The major statistics of single-electron transitions in the present model \textit{in the adiabatic limit} is defined by a short-ranged normal diffusion in the vicinity of the crossing point with a modification caused by anomalous diffusion. We emphasize again that the statistics of electron transitions viewed from the equilibrium ensemble perspective of $F_i^{(ens)}(t)$ is very different. It is primarily determined by anomalous diffusion. The difference is huge! Furthermore, the exact asymptotics for large $t\gg \tau_r$ is given by Eq. (\ref{asymp2}) with $\alpha=0.5$, both in adiabatic and nonadiabatic regimes.

\subsubsection{Short-and-intermediate time statistics in the strictly sub-Ohmic case}

Let us also consider a strictly sub-Ohmic case with $\eta_0=0$. Unfortunately, in this case, there are no simple analytical results available on the whole time scale. However, one can derive a short time asymptotics from Eqs. (\ref{analytic3}), (\ref{approx3}), using the limit $s\to\infty$ and an Abelian theorem. In doing so, we obtain 
\begin{eqnarray}\label{subOhm_ML}
F_i^{(sgl)}(t)\approx E_{1-\alpha/2}\left [-\left( \frac{t}{\tau_{i,sgl}}\right)^{1-\alpha/2} \right ], 
\end{eqnarray}
with 
\begin{eqnarray}\label{tau_single3}
\tau_{i,sgl}\approx \tau_r \left ( \frac{\hbar \sqrt{ \lambda k_BT} }{ 
\sqrt{\pi \Gamma(1+\alpha)/2}\Gamma(1-\alpha/2)\tau_rV_{\rm tun}^2 } \right)^{\frac{2}{2-\alpha}}
\end{eqnarray}
for $\exp(-r_i)\ll 1$, i.e. for a sufficiently large activation energy. This is a very nontrivial result. For $t\ll \tau_{i,sgl}$, it predicts that 
\begin{eqnarray}\label{subOhm_init}
F_i^{(sgl)}(t)\approx \exp \left [-\left(\Gamma t\right)^{1-\alpha/2} \right ], 
\end{eqnarray}
is stretched exponential with a rate parameter 
\begin{eqnarray}\label{Gamma_new}
\Gamma\approx \frac{1}{\tau_r} \left ( \frac{ 
\sqrt{\pi \Gamma(1+\alpha)/2}\tau_rV_{\rm tun}^2 }{\hbar (1-\alpha/2) \sqrt{ \lambda k_BT} } \right)^{\frac{2}{2-\alpha}}\;.
\end{eqnarray}
 This result yields RTDs 
$\psi_i(t)\propto (1-\alpha/2)\exp \left [-\left(\Gamma t\right)^{1-\alpha/2} \right ]/(\Gamma t)^{\alpha/2}$, which for $\Gamma t\ll 1$ agrees with the result by Tang and Marcus in Ref. \cite{TangMarcusPRL}. Furthermore, Eq. (\ref{subOhm_init}) predicts 
$F_i^{(sgl)}(t)\propto 1/(\Gamma t)^{1-\alpha/2}$ for intermediate times 
$\tau_{i,sgl}\ll t\ll \tau_r$, which agrees with the corresponding 
$\psi_i(t)\propto1/(\Gamma t)^{2-\alpha/2}$ obtained by Tang and Marcus  for a Davidson-Cole medium. This prediction is, however, wrong, see below, because of the ultimate failure of non-Markovian Zusman equations. Finally, the same asymptotics (\ref{asymp2}) describes the long-time behavior. This theoretical result is, however, also disproved by numerics based on single trajectories. These two failures signify a significant failure of non-Markovian Zusman equations to describe statistics of single-electron events.

\section{Single trajectory perspective and stochastic simulations}

Now we wish to compare the ensemble perspective based on the generalized Zusman equations with accurate simulations based on a single-trajectory perspective. For this, we perform a Markovian embedding of GLE dynamics (\ref{GLE1}) following a well-established procedure \cite{GoychukPRE09, GoychukACP12}. It allows getting numerical results with a well-controlled numerical accuracy. To this end, the power-law memory kernel, which corresponds to the Caputo fractional derivative, is first approximated by a sum of exponentials, 
\begin{eqnarray}\label{sum}
\eta(t)=\sum_{i=1}^N k_i\exp(-\nu_i t),
\end{eqnarray} 
with the relaxation rates $\nu_i$ and elastic constants $k_i$ obeying a fractal scaling \cite{PalmerPRL85,GoychukPRE09,GoychukACP12},
$\nu_i=\nu_0/b^{i-1}$, $k_i =C_\alpha(b)\eta_\alpha\nu_i^\alpha/\Gamma(1-\alpha)\propto \nu_i^\alpha$. Here, $C_\alpha(b)$ is some constant, which depends on $\alpha$ and a scaling parameter $b$. 
This approximation works well between two memory cutoffs, a short-time cutoff $\tau_l=b/\nu_0$ and a large-time cutoff $\tau_h=b^{N-1}/\nu_0$.
Already for the decade scaling with $b=10$, one arrives at the accuracy of 4\%  (for $\alpha=0.5$, with $C_{0.5}(10)\approx 1.3$). Moreover, with $b=2$ and $C_{0.5}(2)\approx 0.39105$ it can be improved up to $0.01\%$ \cite{MMNP13}, if necessary. 
Next, one introduces a set of auxiliary overdamped Brownian quasi-particles with the coordinates $y_j$.
They are  elastically coupled to the reaction coordinate with coupling constants $k_j$ and are subjected to the viscous friction with the friction coefficients $\eta_j=k_j/\nu_j$ and the corresponding thermal noises related to the friction by the FDT. For the dynamics in the quantum state $i$ we have:
\begin{eqnarray}
  &&\eta_0\dot{x}=-\kappa(x-x_0\delta_{i,2})-\sum_{j=1}^N k_j(x-y_j)+\xi_0(t), \nonumber \\
  &&\eta_j\dot{y}_j=k_j(x-y_j)+\xi_j(t), 
\label{GLEf}
\end{eqnarray}
where $\xi_j(t)$ are $N$ additional uncorrelated white Gaussian noises, $\langle\xi_i(t)\xi_j(t')\rangle=2k_BT\eta_i\delta_{ij}\delta(t-t')$. Notice, that for the model 
with $\eta_0=0$, the first equation in (\ref{GLEf}) yields $x=(\sum_{j=1}^N k_jy_j+\kappa x_0\delta_{i,2})/(\kappa+\sum_{j=1}^Nk_j)$, at all times,
which is used together with the second equation in (\ref{GLEf}) to formulate the corresponding stochastic algorithm. In this work, we numerically deal, however, primarily with the case of $\eta_0\neq 0$.

The dimension  $N+1$  of a Markovian embedding of non-Markovian one-dimensional dynamics is chosen sufficiently large, so that $\tau_h$ exceeds the largest characteristic time of the simulated dynamics, e.g., the largest residence time in a state occurring in the simulations. 
One should mention that the  Prony series expansions \cite{Prony, Hauer, Park99, Schapery99} of power-law memory kernels similar to one we use naturally emerge within a polymer dynamics \cite{Doi}, however, with a different rule in the hierarchy of relaxation rates $\nu_i$. Namely, $\nu_i=\nu_l i^p$ with $k_i=const$ rather than our $\nu_i=\nu_0/b^i$, in terms of some lowest relaxation rate $\nu_l=1/\tau_h$, which yields $\eta(t)\propto 1/t^{1/p}$ between two cutoffs \cite{McKinley}. For example, the Rouse polymer model corresponds to $p=2$ with $\alpha=0.5$ \cite{Doi}. The corresponding Markovian embedding, which would reproduce the results of this paper, would be extraordinarily large, about $10^5$ \cite{PCCP18}. It would be simply not feasible numerically. Nevertheless, this existing relation to polymer models provides a perfect justification of our numerical approach. It is especially well suitable to model anomalous dynamics of the reaction coordinate in proteins. The choice of a particular Markovian embedding is a trade-off between the numerical accuracy and feasibility of simulations, which can run for an extraordinarily long time. [Some simulations run for a month on a standard PC]. This is the reason why we choose  an embedding with $b=10$, rather than $b=2$. With $N=12$, and $\nu_0=10^3$ for $\alpha=0.5$ and $\eta_0=0.1$, this choice allows to arrive at the numerical accuracy of about 5\% in stochastic simulations. 

Simulations of Eq. (\ref{GLEf}) are done using stochastic Heun algorithm with a time step of integration $\delta t$ which we vary from $\delta t=10^{-4}$ (maximal) to $\delta t=10^{-7}$ (minimal) to arrive at reliable results. If the crossing point $x^*$ is met between two subsequent positions of the reaction coordinate, $x_{k+1}$ and $x_k$, a corresponding instant velocity is calculated as $v_k=(x_{k+1}-x_k)/\delta t$, and then one decides if a jump occurs onto the different electronic curve, or not, in accordance with the probability in Eq. (\ref{LZS}). Notice that even if formally $v_T^2=\infty$ within the overdamped model, both the realizations $v_k$ and $\langle v_k^2\rangle$ remain always finite because $\delta t$ is finite. However, a linearization of Eq. (\ref{LZS}), in fact, naturally occurs. Eq. (\ref{LZS}) was used for generality, to avoid an additional approximation. The results are not different from using Eq. (\ref{Landau}) instead of Eq. (\ref{LZS}). The former directly follows from Eq. (\ref{GoldenRule}). Hence it precisely corresponds to the reaction term in Eq. (\ref{Zusman}). We showed earlier in the case of Markovian dynamics using a level-crossing theory, see Eqs. (21)-(25) in Ref. \cite{PCCP17} that this approximation also works upon the inclusion of inertial effects insofar $v_0\ll v_T$, or $|V_{\rm tun}|^2/(\hbar \lambda)\ll v_T/(\pi x_0)$, i.e., tunneling velocity at the crossing point is much smaller than the thermal velocity of the reaction coordinate. Following \cite{PCCP17}, one can establish the same criterion also for the considered non-Markovian case. In the overdamped case considered, the discussed linearization is \textit{implicitly} realized numerically for a sufficiently small $\delta t$. The inclusion of inertial effects in our simulations is straightforward. It can be done using a corresponding Markovian embedding of GLE dynamics from Refs. \cite{GoychukPRE09, GoychukACP12}. However, we reserve it for a separate study that is clearly beyond generalized Zusman equations, which neglect the inertial effects entirely. Then, the use of Eq. (\ref{LZS}) instead of (\ref{Landau}) is generally very essential. 

\subsection{Scaling units and choice of parameters}

Time is scaled  in simulations in the units of Cole-Cole relaxation time $\tau_r$, and the scaled $\eta_0$ corresponds to the initial Debye relaxation time $\tau_0$ in the units of $\tau_r$.  Furthermore, reorganization energy $\lambda$ is scaled in the units of $E_{sc}=\hbar/\tau_{sc}$, Let us fix $\tau_{sc}=2$ ps, which is about Debye relaxation time in the bulk water. Then, $E_{sc}$ is about $2.5\;{\rm cm^{-1}}$ in spectroscopic units. Scaled temperature $k_BT$ will be fixed to $0.1$ of the scaled $\lambda_{sc}=\lambda/E_{sc}$. For room temperature, $k_BT=0.025$ eV, $\lambda \approx 2000\;{\rm cm^{-1}}\approx 0.25$ eV. Such values of $\lambda$ are typical for ET in some proteins, or related molecular structures, e.g. for azurin dimer \cite{MikkelsenPNAS}. Furthermore, the tunnel coupling will be scaled in the units of $\lambda \sqrt{\tau_{sc}/\tau_r}$. In our simulations, $\tau_r$ is an arbitrary parameter and the results can be interpreted for different physical values of $\tau_r$ accordingly. For example, let is take $\tau_r=2\;\mu{\rm s}$. Then, $V_{\rm tun}$ is scaled in units of $10^{-3}\lambda=2.5\times 10^{-4}$ eV and $V_{\rm tun}=0.01$ would correspond to $V_{\rm tun}=2.5\times 10^{-6}$ eV. Such small tunnel couplings are often met in protein structures, e.g., in azurin dimers \cite{MikkelsenPNAS}. Given the results of this work, an adiabatic non-exponential ET regime can occur in a Cole-Cole medium for rather small tunnel couplings, which should be a great surprise for many ET researchers. We do not mean, however, any particular case of ET in proteins. Our consideration is generic, and the readers can play with the parameters $\tau_r$ and $V_{\rm tun}$. For example, if to assume
$\tau_r=200\;{\rm ps}$, then $V_{\rm tun}$ is scaled in the units of $0.025$ eV or thermal energy at a room temperature. Then, physical $V_{\rm tun}=2.5\times 10^{-6}$ eV would correspond to $V_{\rm tun}=10^{-4}$ scaled, which in turn would correspond to a non-adiabatic ET, see below.

ET literature discusses a possible influence of medium relaxation on ET transfer rates assuming $\tau_r$ be varying in the range from pico- to nanoseconds, in the case of Debye media, see, e.g., in reviews \cite{GrayPNAS, Skourtis10} and the references cited therein. However, electron tunneling coupling in molecular compounds displaying solvent dynamical effects, see, e.g., in \cite{Chakrabarti09}, can also be significantly larger, up to room $k_BT$.
In numerics, we considered the symmetric case of $\epsilon_0=0$, with activation barriers $r_1=r_2=r=\lambda/(4k_BT)=2.5$ in the scaled units, like for azurin dimer. In the scaled units, non-adiabatic rates read $k_{1,2}^{(na)}=\sqrt{\pi}V_{\rm tun}^2\lambda e^{-r}/\sqrt{T}$.

\section{Results and Discussion}

\subsection{Population dynamics}

\begin{figure}
\vspace{1cm}
\resizebox{0.95\columnwidth}{!}{\includegraphics{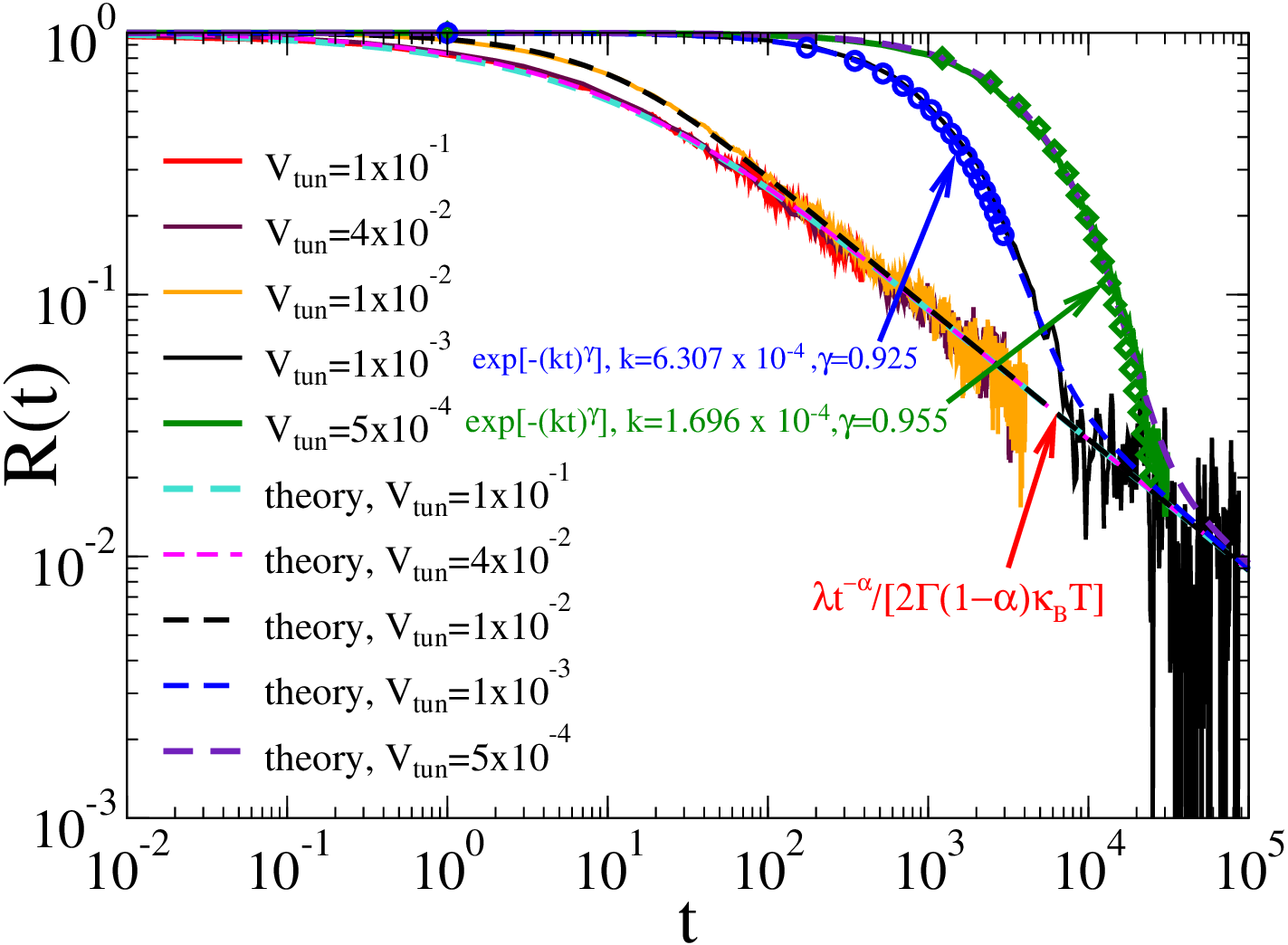}}
\caption{(color online) Relaxation of electronic states populations \textit{vs.} time scaled in the Cole-Cole relaxation constant $\tau_r$, for 5 different values of tunneling coupling scaled in the units of $\lambda\sqrt{\tau_{sc}/\tau_r}$, for a symmetric ET, $\epsilon_0=0$, with $\lambda=800$ in the scaled units of $E_{sc}=\hbar/\tau_{sc}$ and $k_BT=0.1\lambda$. For $\tau_{sc}=2$ ps, $\lambda=0.25$ eV. Adiabatic time 
const $\tau_{ad}=25$. Full lines depict the numerical results obtained from many-trajectory simulations with $10^4$ particles. The dashed lines correspond to the analytical result in Eq. (\ref{Rexact}) from the generalized Zusman equations. The agreement is remarkable indeed! The symbols correspond to stretched-exponential fits of some numerical results with the parameters shown in the plot.
The correspondingly scaled total nonadiabatic rate $k_{na}$ is $k_{na}\approx 7.3614$  for
$V_{\rm tun}=10^{-1}$. Furthermore, $k_{na}\approx 1.7782$ for $V_{\rm tun}=0.04$, $k_{na}\approx 0.07361$
for $V_{\rm tun}=0.01$, $k_{na}\approx 7.361\times 10^{-4}$
for $V_{\rm tun}=10^{-3}$, and $k_{na}\approx 1.840\times 10^{-4}$
for $V_{\rm tun}=5\times 10^{-4}$. With $\kappa_{ad}\approx 184.034$ for $V_{\rm tun}=0.1$,
and $\kappa_{ad}\approx 29.445$ for $V_{\rm tun}=0.04$, anomalous transport is clearly adiabatic for these parameters, as well as for all larger tunnel couplings. For $V_{\rm tun}=0.01$, $\kappa_{ad}\approx 1.840$ and ET is still near to adiabatic. For $V_{\rm tun}=10^{-3}$, $\kappa_{ad}\approx 1.84\times 10^{-2}$, and 
for $V_{\rm tun}=5\times 10^{-4}$, $\kappa_{ad}\approx 4.60\times 10^{-3}$, which is the case of anomalous nonadiabatic ET featured by a power-law heavy tail, and a stretched exponential main course.}
\label{Fig1}       
\end{figure}

First, we studied the dynamics of populations numerically. For this, $M=10^4$ particles were propagated, all started in one electronic state, with the reaction coordinate initially equilibrated, and each making huge many transitions between two electronic states during the relaxation process. The resulting relaxation function $R(t)$ is shown in Fig. \ref{Fig1}, for five different values of scaled $V_{\rm tun}$. Two cases correspond to anomalous adiabatic ET, as explained in Fig. \ref{Fig1} caption. Two others correspond to anomalous non-adiabatic regime, and one to an intermediate case close to the adiabatic regime. Notice a remarkable agreement between the theory based on the generalized Zusman equations, namely the result in Eq. (\ref{Rexact}), and the trajectory simulations both in the adiabatic regime and nonadiabatic regime. Deep adiabatic regime starts already from $V_{\rm tun}=0.04$ in Fig. \ref{Fig1}, which corresponds e.g. to $V_{\rm tun}=1\times 10^{-5}$ eV for $\tau_r=2\;\mu{\rm s}$ and $\tau_{sc}=2$ ps, in physical units.  Even for $\tau_r=50$ ps (a typical value for fractional protein dynamics, which can be attributed to fractons), the corresponding value $V_{\rm tun}=2\times 10^{-3}$ eV is pretty small. Such small values of $V_{\rm tun}$ indicate that the medium dynamics can enslave ET, in the ensemble sense, and make it adiabatic even for very small tunnel couplings. In such a deeply adiabatic regime, anomalous ET is well described by a simple dependence $R(t)\approx E_{1/2}\left [-\sqrt{\frac{t}{\tau_{ad}}}\right ]=e^{t/\tau_{ad}}{\rm erfc}\left [ \sqrt{\frac{t}{\tau_{ad}}}\right ]$. Notice that one cannot define here a proper adiabatic rate, and the quantity $\gamma_{1/2}=1/\sqrt{\tau_{ad}}$ can be interpreted as a fractional adiabatic ET rate of the fractional order $1/2$. Initially, for $t\ll \tau_{ad}=4\tau_r[E^{(a)}/k_BT]^2=\tau_r \lambda^2/(2k_BT)^2$, $R(t)\approx \exp[-\sqrt{2t/\pi\tau_{ad}}]$ is stretched exponential.  For $t\gg \tau_{ad}$, a power law tail emerges, $R(t)\sim (\lambda/2k_BT)/\sqrt{\pi t}$. The agreement with the theory implies that the result in Eq. (\ref{puzzle}) is universally valid also for other values of $\alpha$, and $\eta_0$, including $\eta_0=0$. Also in the non-adiabatic ET regime this universal behavior is seen in Fig. \ref{Fig1}, even if it becomes buried in the population fluctuations due to a finite $M$ (mesoscopic noise) with diminishing $V_{\rm tun}$. Indeed, for $R(t)\sim 1/\sqrt{M}$ and below, which is $0.01$ or 1\% for $M=10^4$ in Fig. \ref{Fig1}, the relaxation becomes masked by the population fluctuations.  Likewise, this feature may be blurred by noise also in real experiments. It is expected to be a universal feature of sub-Ohmic incoherent dynamics. The analytical result in Eq. (\ref{Lambert}) predicts for $V_{\rm tun}=5\times 10^{-4}$ in Fig. \ref{Fig1} that the crossover time $t_c$ to the power law behavior is $t_c\approx 2.146\times 10^4 $, and the corresponding $R(t_c)\approx 0.0193 $, i.e. nearly 2\% of the rest population relaxation follows a universal power law. It agrees with numerics fairly good in Fig. \ref{Fig1}. However, a good agreement with the theoretical values of $t_c$ and $R(t_c)$ is not expected for larger tunnel couplings because then the major kinetics deviates strongly from a single-exponential.  It is rather stretched exponential, see in Fig. \ref{Fig1} for $V_{\rm tun}=1\times 10^{-3}$. One can regard the power exponent $\gamma$ of stretched-exponential larger than $0.95$ as one close to $\gamma=1$ of single exponential. Then, Eq. (\ref{Lambert}) is expected to work.  The value $\gamma\approx 0.925$ for $V_{\rm tun}=1\times 10^{-3}$ is not that close. The importance of the analytical result in Eq. (\ref{Lambert}) lies in the fact that it allows to predict correctly $t_c$ and the weight $R(t_c)$ of the power law relaxation tail for such small $V_{\rm tun}$, which are not attainable for numerical analysis. For example, to obtain the relaxation curve for $V_{\rm tun}=5\times 10^{-4}$ in Fig. \ref{Fig1}, it took more than one month of the computational time on a standard modern PC. The numerics are hardly feasible on standard PCs already for $V_{\rm tun}=1\times 10^{-4}$, with the same numerical accuracy.

\subsection{Electronic transitions from the equilibrium ensemble perspective}

\begin{figure*}
\includegraphics[width=8cm]{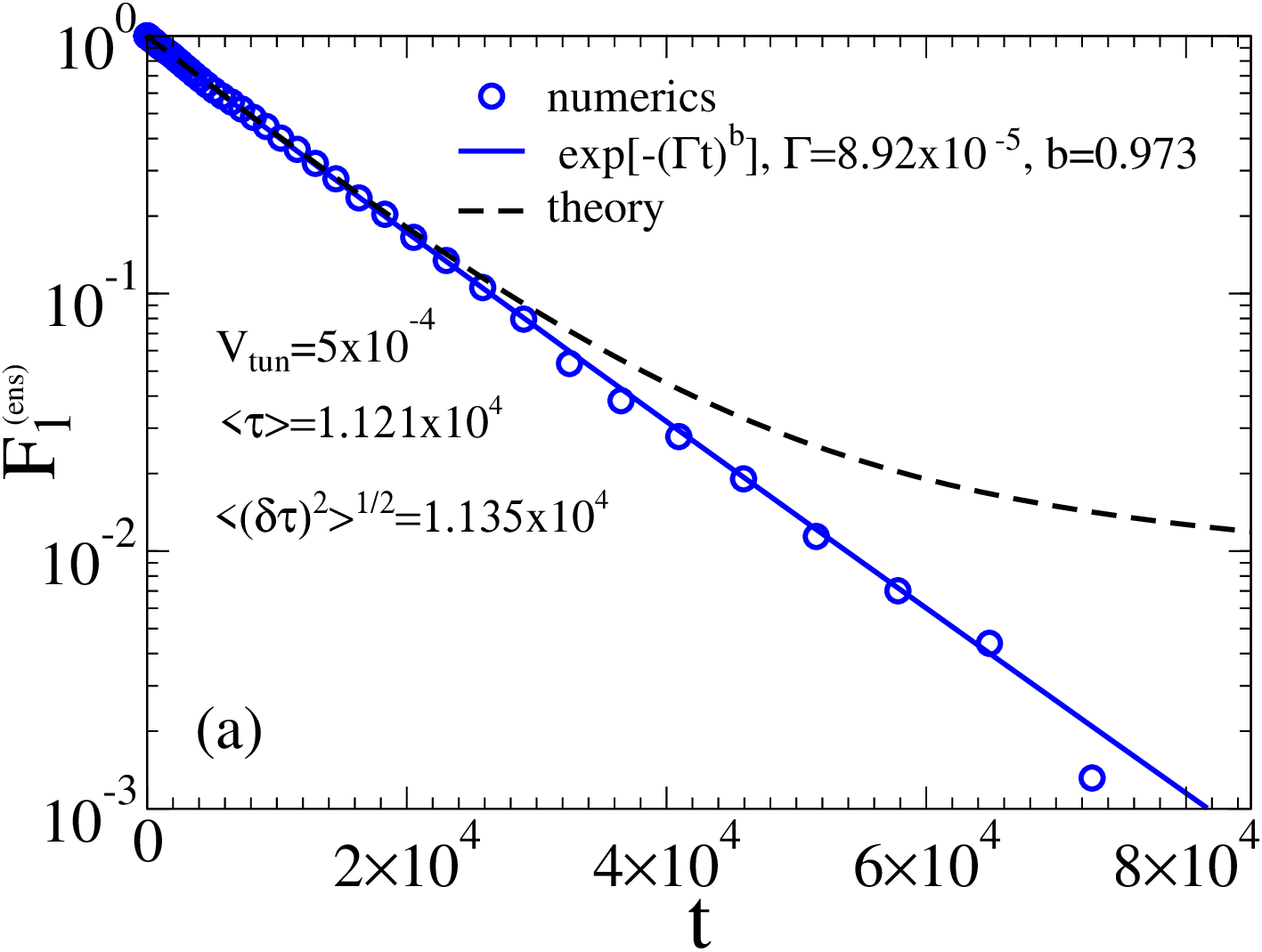}
\hspace*{0.8cm}
\includegraphics[width=8cm]{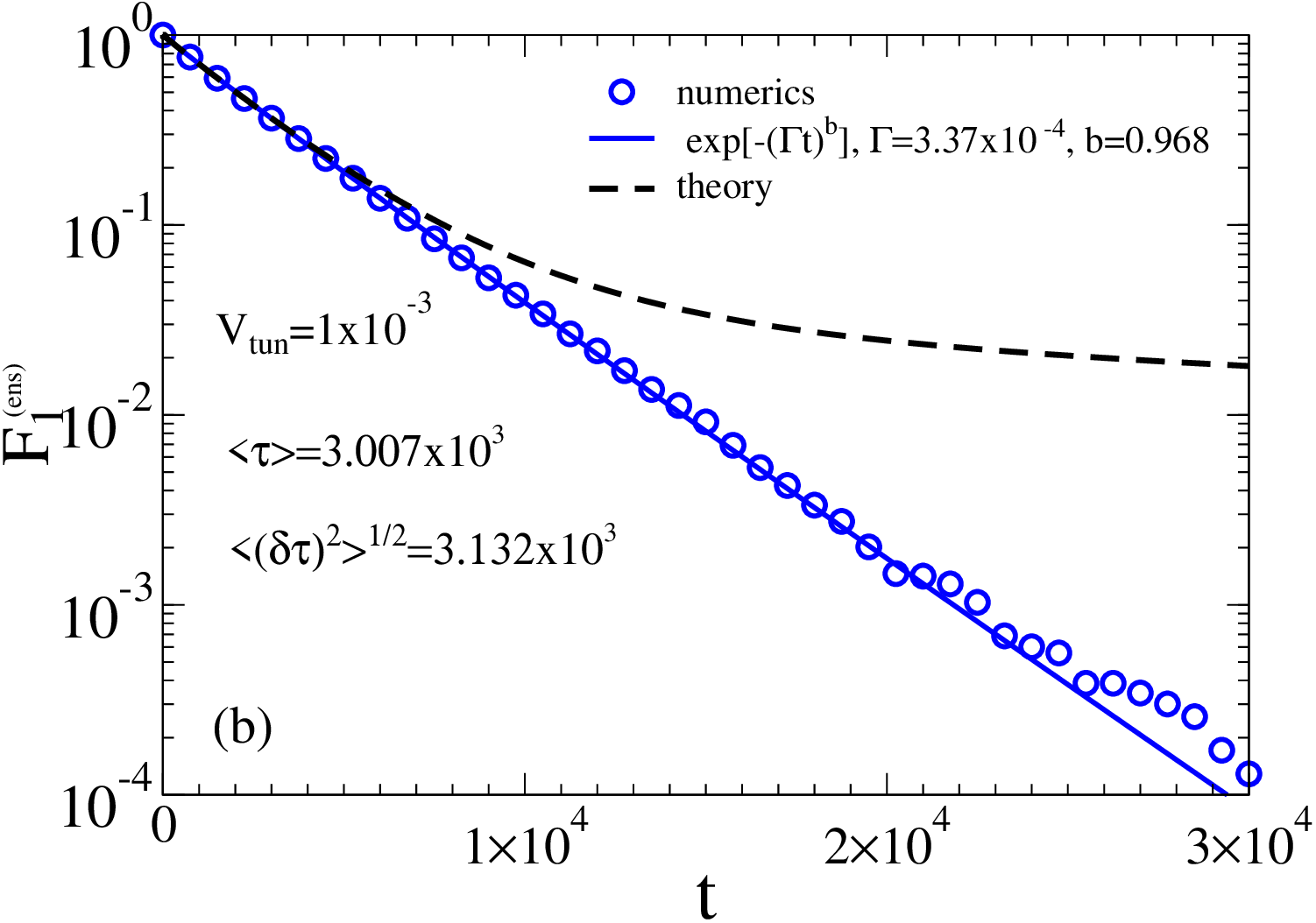}\\[0.8cm]
\includegraphics[width=8cm]{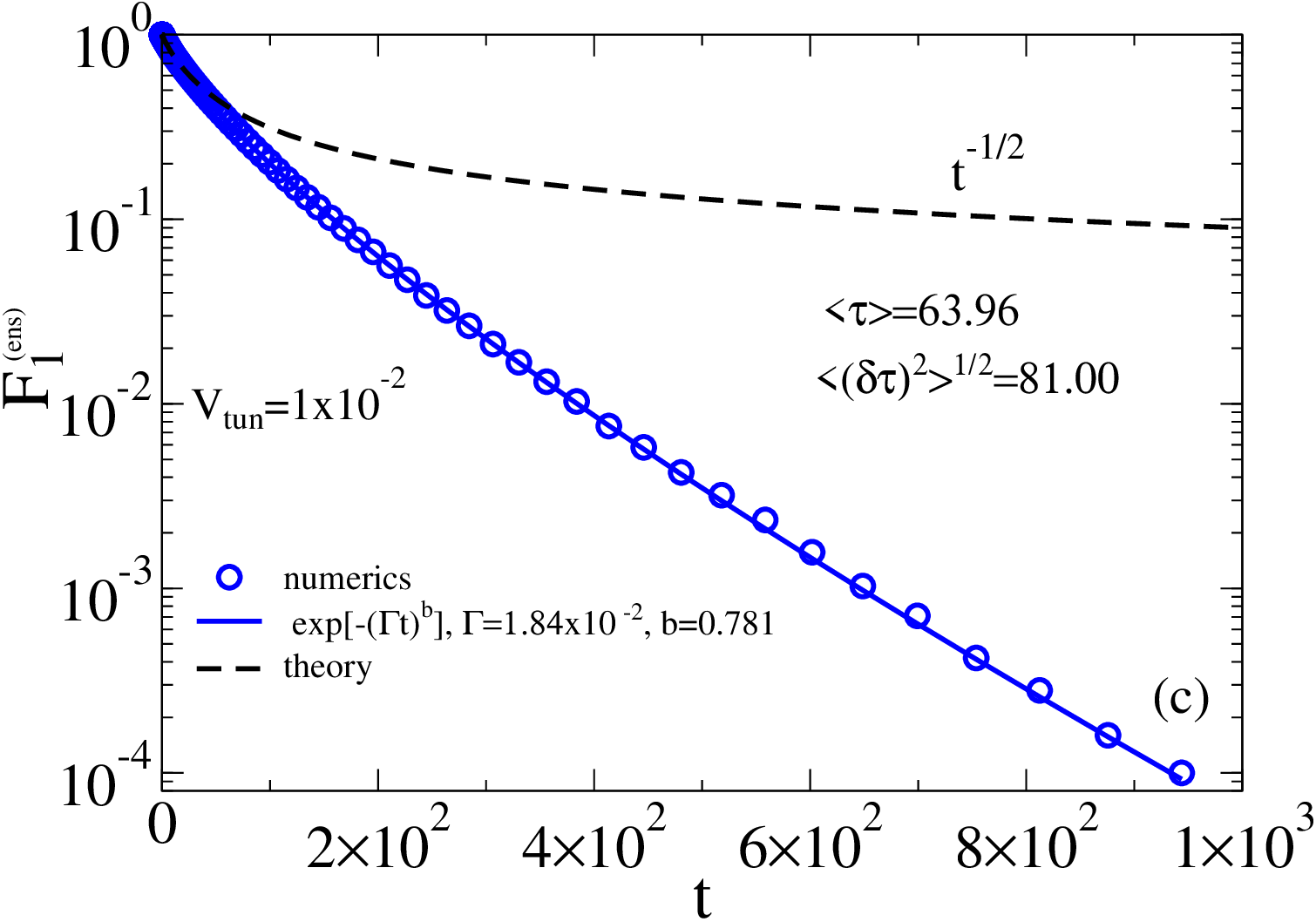}
\hspace*{0.8cm}
\includegraphics[width=8cm]{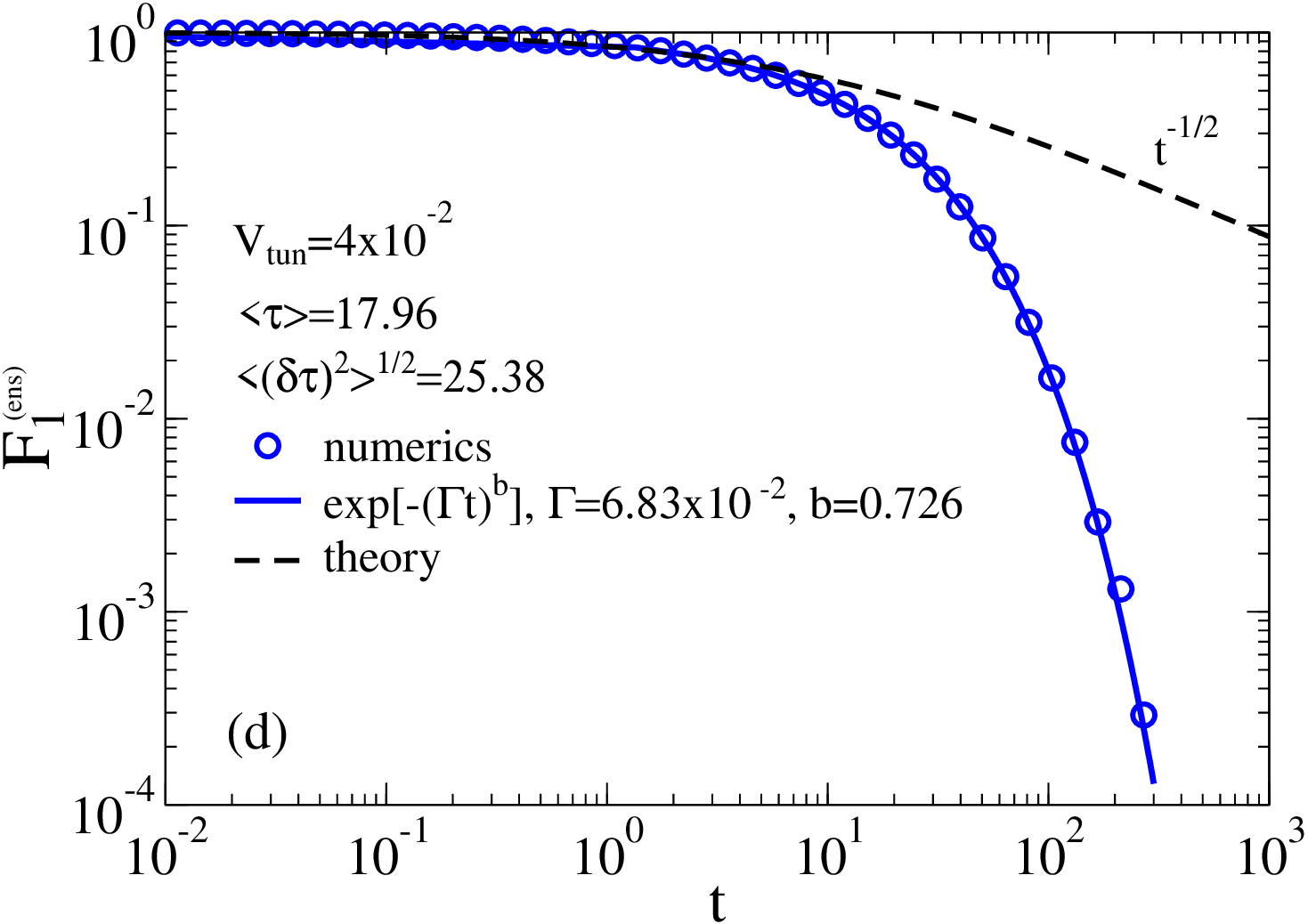}\\[0.8cm]
\includegraphics[width=8cm]{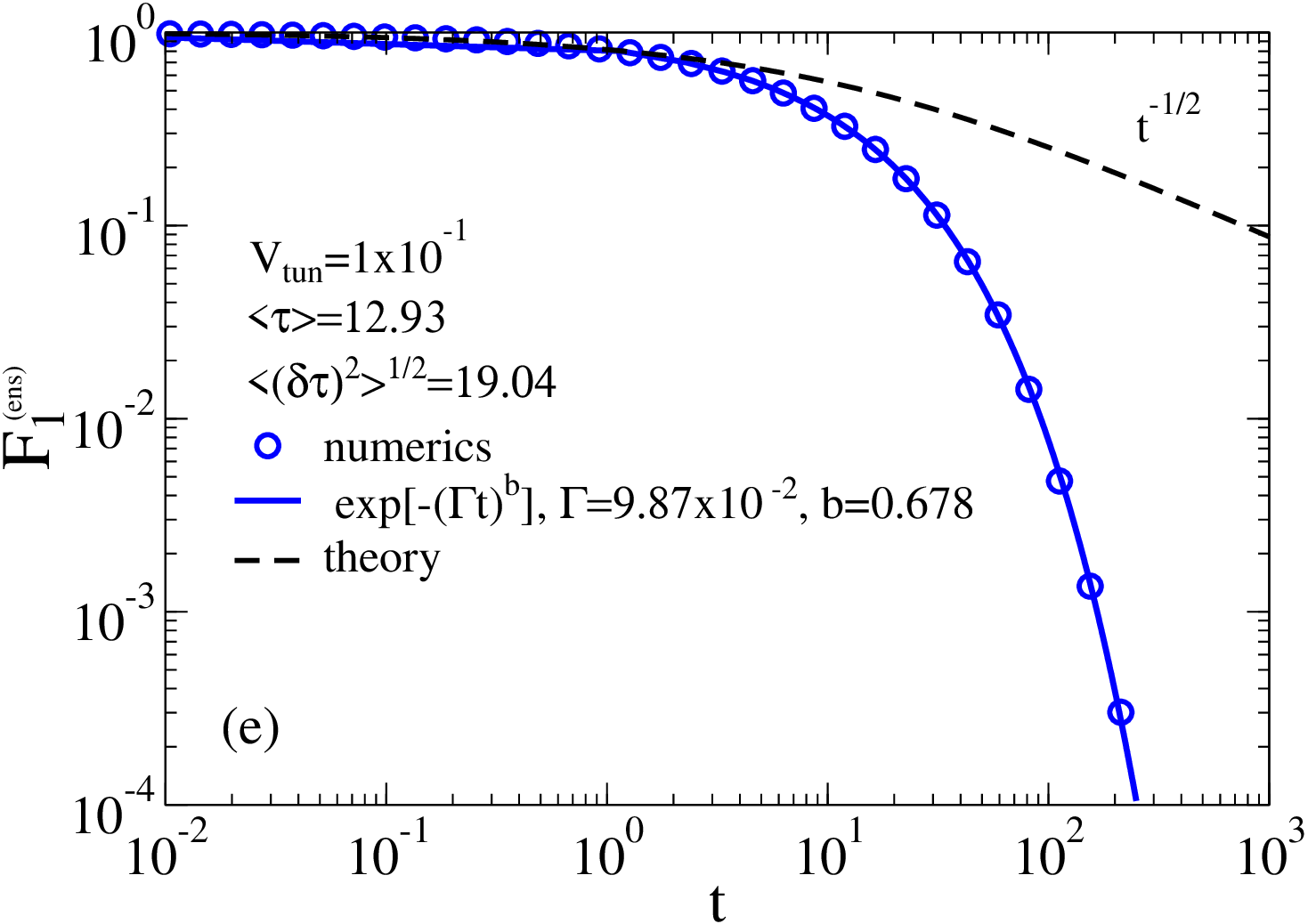}
\hspace*{0.8cm}
\includegraphics[width=8cm]{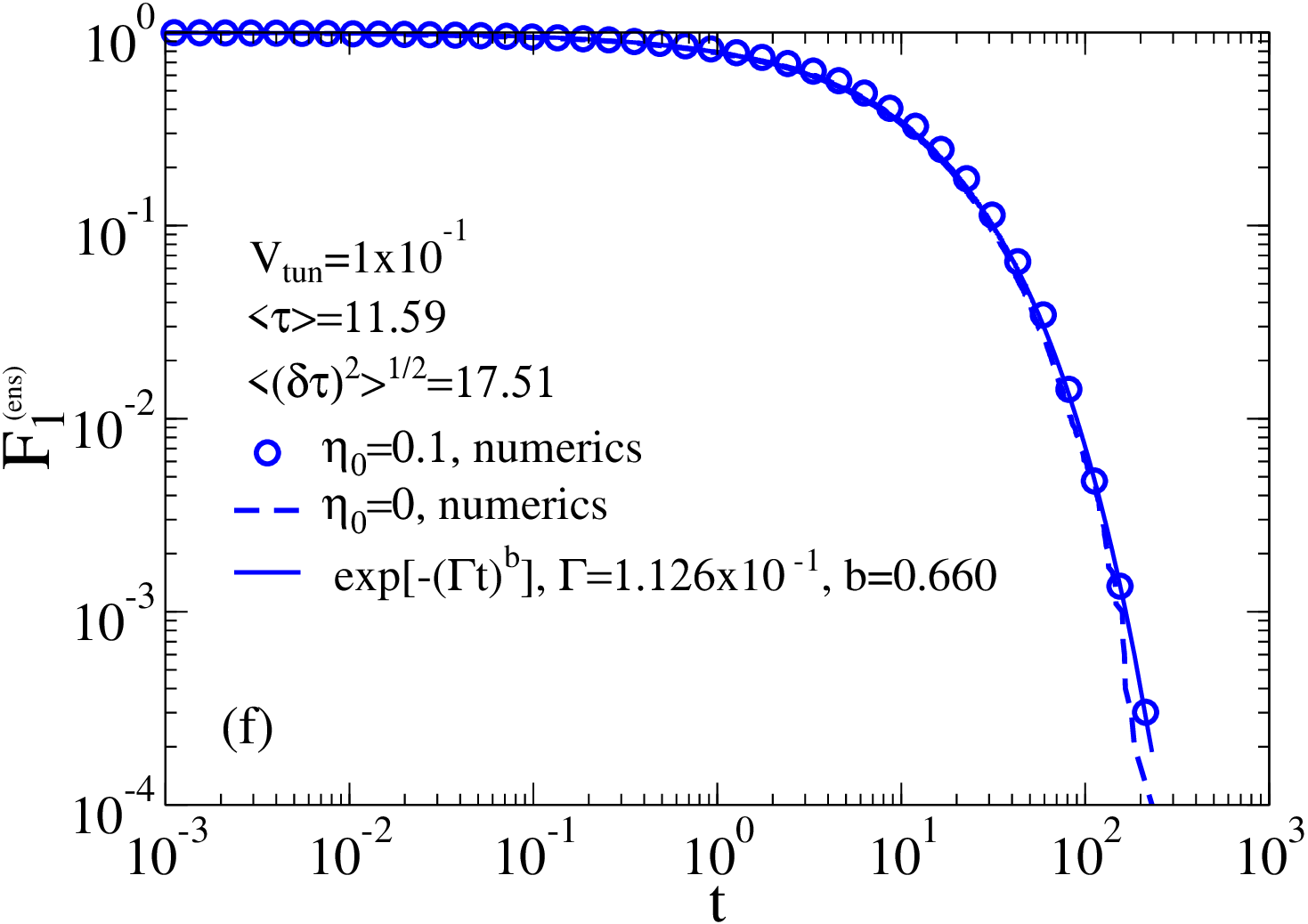}
\caption{(Color online) Time-decay of the survival probability in the first state calculated from the trajectory simulations done using a thermal equilibrium preparation of the reaction coordinate at the initial time. Time is scaled in the units of $\tau_r$. Any single trajectory is terminated once a jump into another electronic states occurs. Statistics is derived from $10^4$ trajectories. The dashed black line depicts the theory result from Eqs. 
(\ref{analytic2}), (\ref{Rexact}), as described in Sec. III,B. Notice, that by a sharp contrast with Fig. \ref{Fig1}, where a related result agrees with numerics very well, for the survival probabilities it fails completely. First, not only the mean residence time is finite (the theory predicts that it is infinite), but also the variance of RTD is finite. The corresponding numerical values of mean values and dispersion coefficients are given in different panels for different values of $V_{\rm tun}$ shown therein. Second, the theoretical tail prediction, $F_i(t)\propto 1/t^{1/2}$, is completely wrong. Survival probability is well described by a stretched exponential, which tends to a single exponential with the rate given by the MLD rate $k_1^{(na)}$ in the limit $V_{\rm tun}\to 0$, see the main text for more detail.
In the panels (a)-(e), $\eta_0=0.1$, whereas in (f) also a strictly sub-Ohmic case of $\eta_0=0$ is compared with the case of $\eta_0=0.1$ in the panel (e). This comparison does not reveal a statistically significant difference. Thus, a finite but small value of $\eta_0$ only weakly influences survival probabilities from the equilibrium ensemble perspective. }
\label{Fig2}
\end{figure*}

The next important question we address is: What is the survival probability of electrons in an electronic state from the equilibrium ensemble perspective? To answer this question, we prepare all the electrons in one state at the equilibrated reaction coordinate (a different value is taken randomly from the Boltzmann distribution for each electron in the ensemble) and take out an electron from the ensemble once it jumps into another state at the crossing point. The numerical results are depicted in Fig. \ref{Fig2} in comparison with the theoretical results based on the generalized Zusman equations. The theory fails in a very spectacular fashion. First, the mean residence time in the state is finite, at odds with the theory predicting infinite MRT. Also, the variance of RTD is finite. Second, the power law tail, $F_1^{(ens)}(t)\propto 1/\sqrt{t}$, which the theory predicts, is absent. Instead, the survival probability is well described by a stretched exponential dependence, $F_1^{(ens)}(t)\approx \exp[-(\Gamma t)^b]$, in some transient parts, or even for all times. A similar failure on the non-Markovian FPE to describe the statistics of subdiffusive transitions in bistable dynamics has already been described earlier \cite{GoychukPRE09,GoychukACP12}, and the related fiasco of the non-Markovian generalization of Zusman theory is explained below.  No doubts, in the strict non-adiabatic limit of $V_{\rm tun}\to 0$, survival probabilities  are strictly exponential, $F_i^{(ens)}(t)=\exp(-k^{(na)}_i t)$, with non-adiabatic MLD rates. Already, for the smallest $V_{\rm tun}=5\times 10^{-4}$, $b\approx 0.973$, see in Fig. \ref{Fig1}, a, and $\Gamma\approx 8.92\times 10^{-5}$, which is not much different from the corresponding MLD rate $k^{(na)}_1\approx 9.20 \times 10^{-5}$. The numerical $\langle \tau \rangle \approx 1.121\times 10^{4}$ also does not differ much in this non-adiabatic regime from $1/k^{(na)}_1\approx 1.087\times 10^{4}$. With increasing tunnel coupling, $b$ becomes smaller. For $V_{\rm tun}=1\times 10^{-3}$, $b\approx 0.892$ initially (not shown) and $b\approx 0.968$ for large times with $\Gamma\approx 3.37\times 10^{-4}$, which still does not differ much from the corresponding $k^{(na)}_1\approx 3.68 \times 10^{-4}$, see in Fig. \ref{Fig2}, b. Also, numerical $\langle \tau_1 \rangle \approx 3.007\times 10^{3}$ is only slightly larger,  due to adiabatic corrections, than $1/k^{(na)}_1\approx 2.717\times 10^{3}$. This is still a non-adiabatic ET regime. The smallest value $b\approx 0.678$ is arrived for the largest $V_{\rm tun}=0.1$ in our simulations, see in Fig. \ref{Fig2}, e. In this case,  $\langle \tau \rangle \approx 12.93$, which is essentially larger than $1/k^{(na)}_1\approx 0.272$. It can be regarded as an effective inverse adiabatic rate, which is essentially smaller than  $k^{(na)}_1$. There is no any signature of a power law behavior also in this case. In Fig. \ref{Fig2}, f, we plotted also the survival probability for the strict Ohmic case of $\eta_0=0$, in comparison with the corresponding result for $\eta_0=0.1$. The comparison shows that the discrepancy between two cases on the ensemble level is almost negligible. For smaller $V_{\rm tun}$, such a discrepancy is expected to be even smaller.

\subsection{Electronic transitions from single trajectories}

\begin{figure*}
\includegraphics[width=8cm]{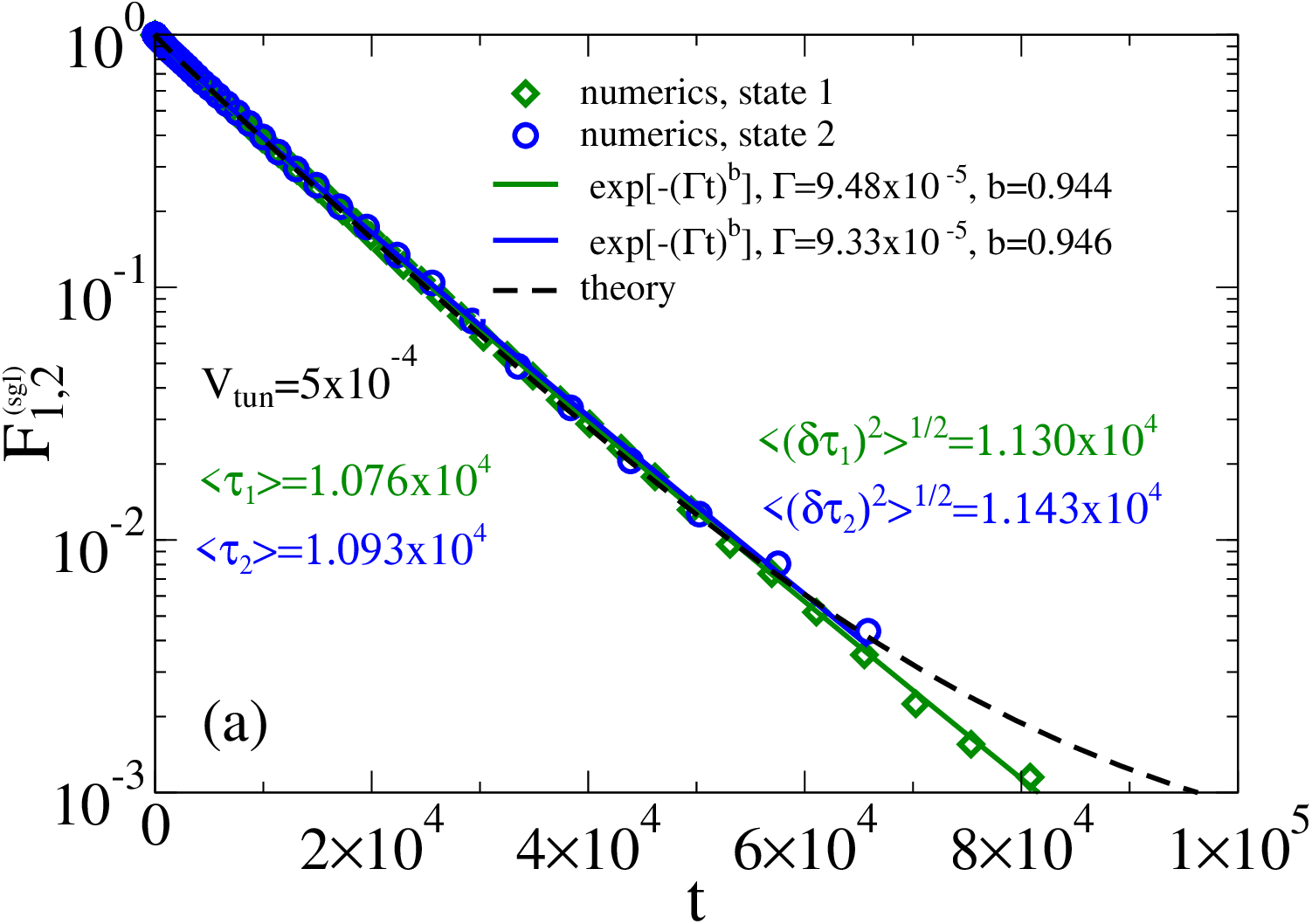}
\hspace*{0.8cm}
\includegraphics[width=8cm]{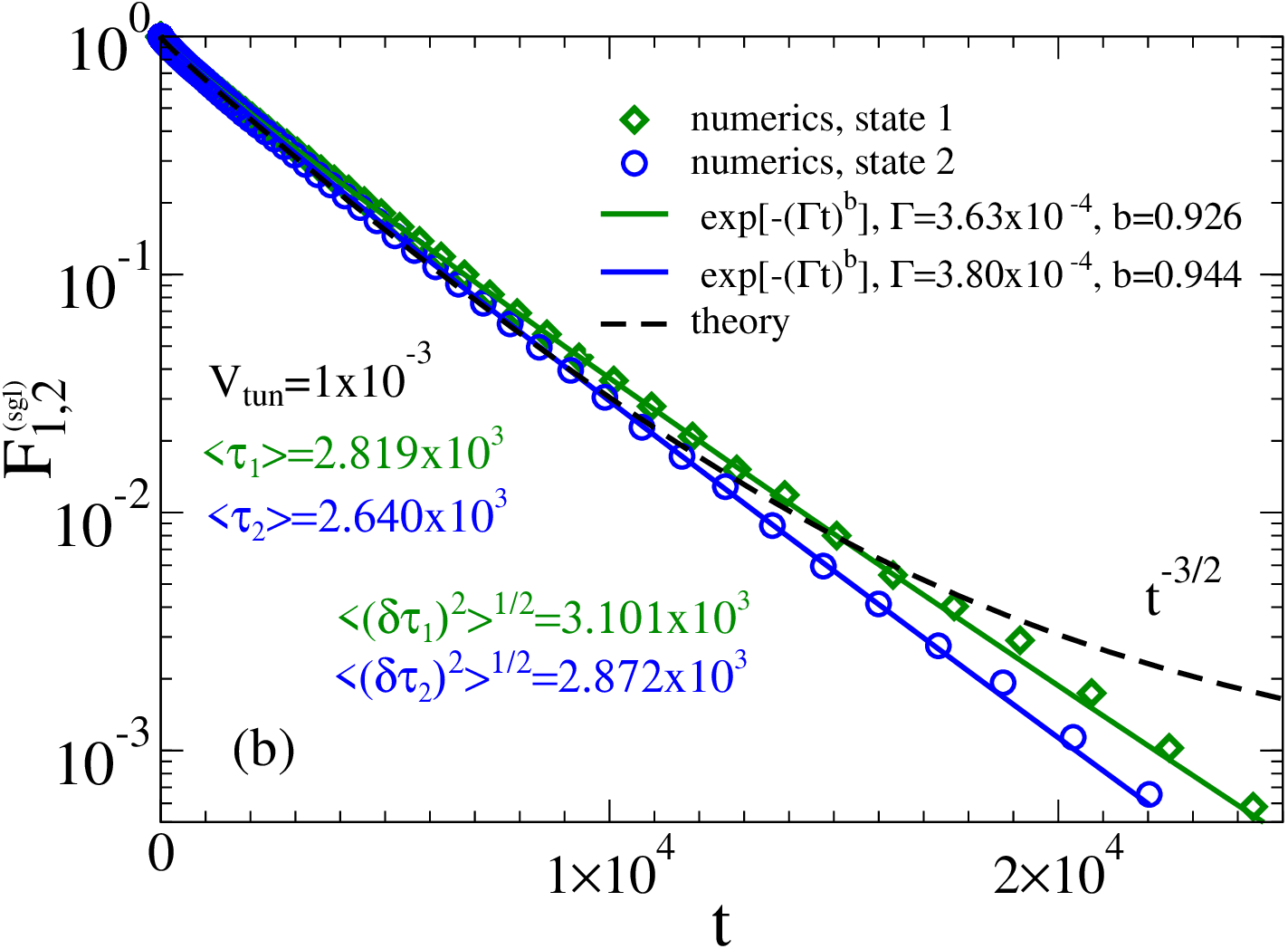}\\[0.8cm]
\includegraphics[width=8cm]{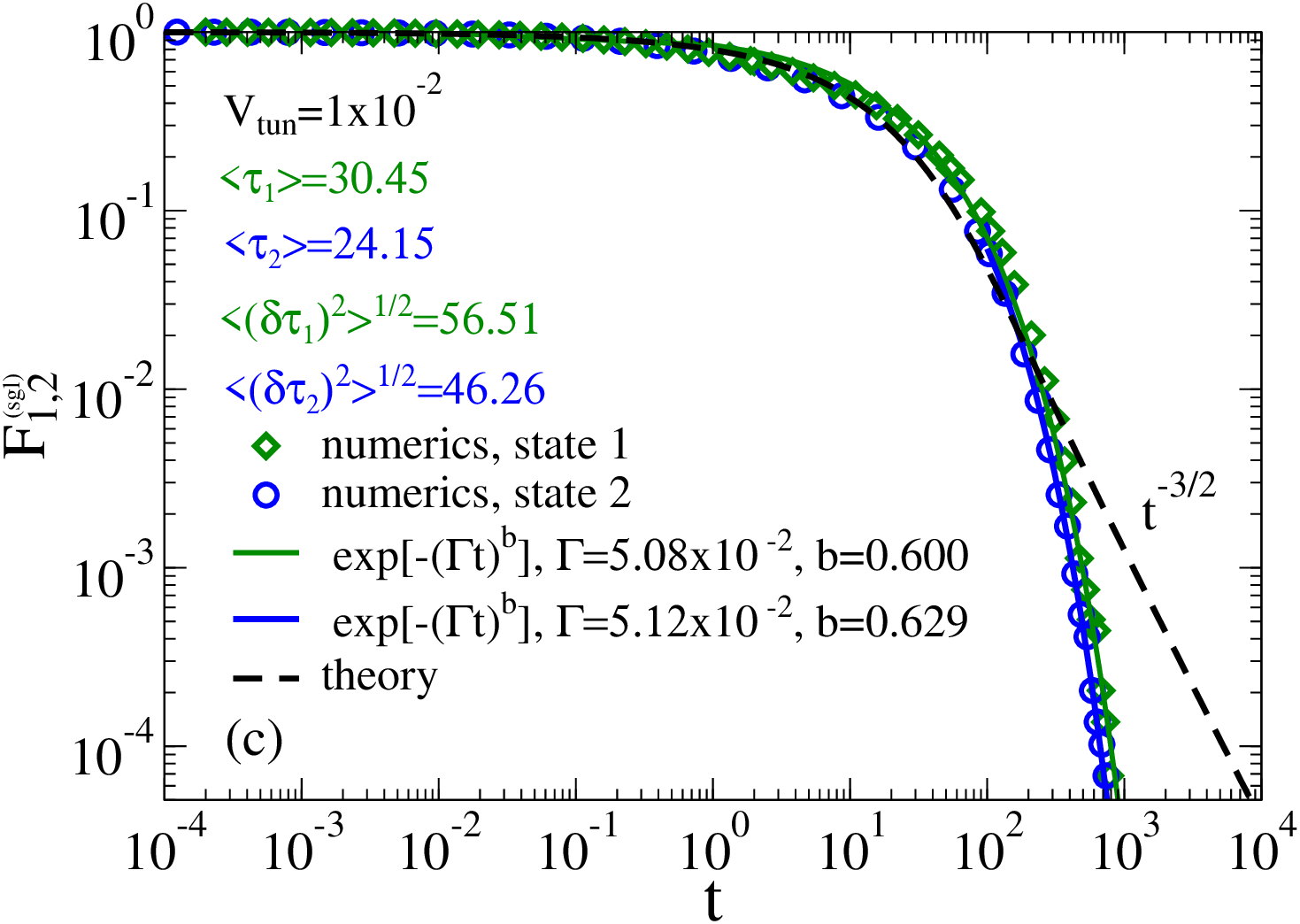}
\hspace*{0.8cm}
\includegraphics[width=8cm]{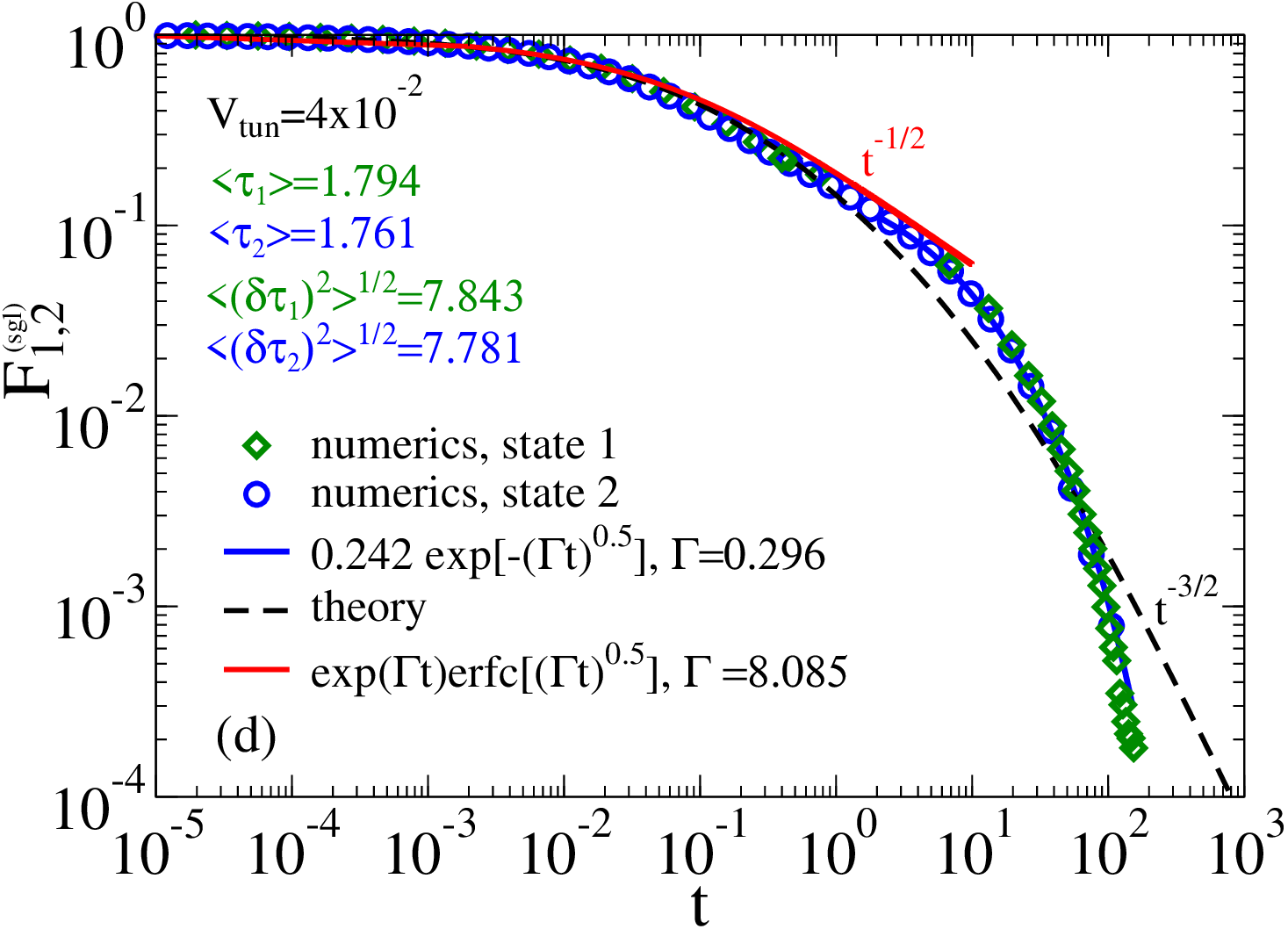}\\[0.8cm]
\includegraphics[width=8cm]{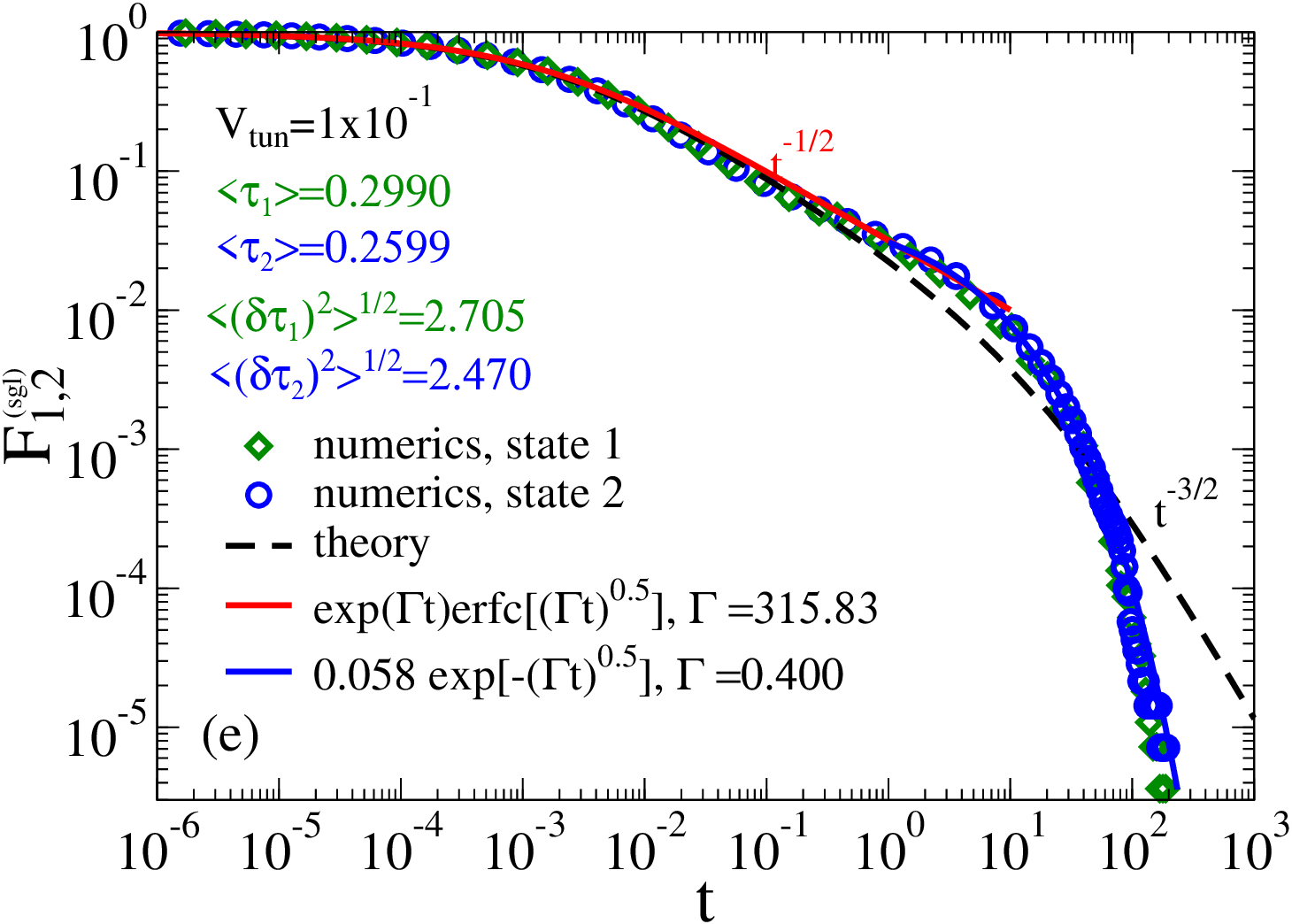}
\hspace*{0.8cm}
\includegraphics[width=8cm]{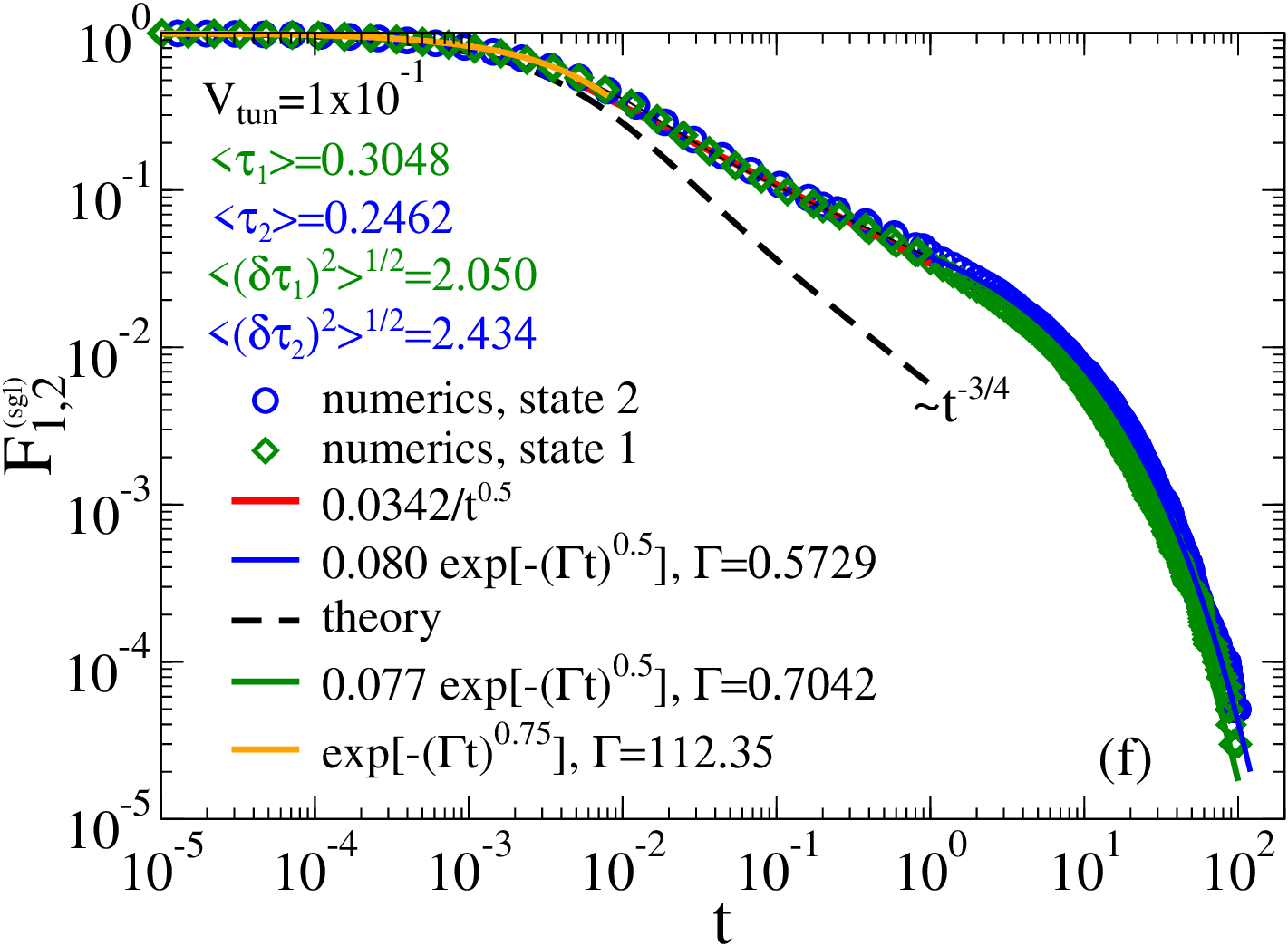}
\caption{(Color online) Survival probabilities in two states \textit{vs.} time (in units of $\tau_r$) from a single trajectory perspective. The numerical data are shown by symbols and their various fits (with the parameters shown in the plots) by the full lines. The results of the analytical theory based on the generalized Zusman equations in the contact approximation are depicted by the dashed black lines. In the panels (a)-(e), $\eta_0=0.1$. In the panel (f), $\eta_0=0$. The values of the tunnel coupling $V_{\rm tun}$ are shown in the corresponding panels. }
\label{Fig3}
\end{figure*}

To derive statistics from single trajectories, a very long single trajectory is stochastically propagated and the residence time distributions in both electronic states are derived from the pertinent numerical experiments, like in Ref. \cite{PCCP17}. The results are shown in Fig. \ref{Fig3}. For the smallest $V_{\rm tun}=5\times 10^{-4}$ in such experiments, see in part (a), the theoretical result in Eq. (\ref{Fexact}) agrees with numerics pretty well up to $F_i^{(sgl)}(t)\approx 0.004$, i.e., it describes almost 99.6 \% of the decay of the survival probability, which is a remarkable success of the theory based on generalized Zusman equations. The survival probability is approximately stretched exponential on the whole time scale. The statistical discrepancy between the left and right state distributions because of a finite sample size is minimal. The mean stretched-exponential power exponent $b\approx 0.945$ is smaller than $b\approx 0.973$ in Fig. \ref{Fig2}, a. However, there are no doubts that the both exponents will approach unity (a strictly exponential distribution) with a further diminishing $V_{\rm tun}$. Nevertheless, the theory predicts a very wrong power-law tail, which is disproved by numerics. This failure becomes ever more visible with the increase of $V_{\rm tun}$.

In the strictly non-adiabatic limit,  ET is by and large ergodic. The MLD rate well describes it. However, some deviations from single-exponential transfer kinetics and ergodicity become visible even for the smallest tunnel coupling in this paper, which is very different from the Markovian Debye case \cite{PCCP17}. One should emphasize this striking feature once more:  Even if sub-Ohmic ET is strictly exponential and ergodic in the strict non-adiabatic limit judging from the survival probabilities in the electronic states, the relaxation of electronic populations follows asymptotically a power law, as described above. It can, however, be tough to detect due to pure statistics in real experiments. For $V_{\rm tun}=5\times 10^{-4}$, the averaged numerical MRT in an electronic state is $\langle \tau\rangle_{\rm emp}=(\langle \tau_1\rangle+\langle \tau_2\rangle)/2\approx 1.0845\times 10^4$. It nicely agrees with the theoretical prediction $\langle \tau\rangle=1/k_{1,2}^{(na)}\approx 1.0868\times 10^4$.
The theoretical prediction of a power-law tail $F_i^{(sgl)}(t)\sim t^{-3/2}$ is, however, once again, completely wrong.

Next, for  $V_{\rm tun}=1\times 10^{-3}$ in Fig. \ref{Fig3}, b, the survival probabilities in two states are somewhat different. It is unclear why statistics is visible poorer in this particular case, what caused that discrepancy. In fact, the results presented in part (a) are based on $2\times 7583$ electronic transitions, while in part (b) on $2\times 20603$ such transitions. A further increase in the number of transitions would smear out the discrepancy in the part (b). However, it would require a much longer computational time. Nevertheless, the averaged $\langle \tau\rangle_{\rm emp}=(\langle \tau_1\rangle+\langle \tau_2\rangle)/2\approx 2.7395\times 10^3$ agrees nicely with the theoretical $\langle \tau\rangle=1/k_{1,2}^{(na)}\approx 2.7169\times 10^3$. As expected, the power of stretched exponential is smaller. It has the mean value $b\approx 0.935$. Here, the discrepancy with the theory result (especially, concerning the tail of the distribution) becomes stronger. Nevertheless, the theoretical result describes very well about 98\% of the survival probability decay in the state 2.  With a transition to the adiabatic regime, the agreement between the theory and numerics becomes worse, see in part (c) of Fig. \ref{Fig3}. However, in the adiabatic regime, it improves again. Accordingly, in the parts (d) and (e) the theory describes even about 90\% and 98\% of the initial decay, correspondingly. This success is because, in this case, the normal diffusion dominates on the corresponding time scale for the studied case of $\eta_0\neq 0$. The theoretical prediction of a power law tail is, however, wrong, completely. Interestingly, the results in part (c) are derived based on $2\times 29254$ transitions, the statistical discrepancy between distribution in both states is, however, much smaller than in part (b), with a similar number of transitions. In the parts (d) and (e), the discussed numerical asymmetry is also pretty small. However, in those two cases, the samples were much larger, $2\times 80276$ in (d), and $2\times 139866$ in (e). In the last two cases of a well-developed adiabatic regime, the initial decay is well reproduced by the Mittag-Leffler distribution $F_i^{(sgl)}(t)\approx E_{1/2}[-\sqrt{\Gamma t}]$, with $\Gamma=1/\tau_{sgl}\approx 8.085$ in the part (d) and $\Gamma=1/\tau_{sgl}\approx 315.83$ in the part (e), with $\tau_{sgl}$ given by Eq. (\ref{tau_single2}). This agreement is a remarkable success of the theory. Also the mean residence time $\langle \tau_{1,2} \rangle$ in all cases was nicely reproduced by the inverse MLD rate, as the theory predicts. However, the tail of distribution in the well-developed adiabatic regime is always $c_1\exp(-\sqrt{\Gamma t })$, with some weight $c_1$ and rate parameter $\Gamma$, very differently from the power law $t^{-3/2}$, which the theory predicts. Here, the theory fails. Needless to say that in the adiabatic regime survival probabilities viewed from the equilibrium ensemble perspective and the view of single trajectories are completely different. They are characterized by entirely different mean residence times and dispersion, compare with Fig. \ref{Fig2}! Hence, from the kinetic point of view, the electron transfer is non-ergodic in this regime. 

We emphasize, however, that that success of the theory in describing the statistics of single trajectories should not be overestimated. This success in the deep adiabatic regime is because on the corresponding time scale the diffusion is normal. This peculiarity is the reason why the corresponding results are very similar in the main initial part of the corresponding distributions, apart from a very different tail, to the results obtained within the normal diffusion Zusman equations, see in Ref. \cite{PCCP17}. Here, a new profound non-ergodic feature is manifested. Namely, in sharp contrast with this normal diffusion feature on the level of single trajectories, the statistics of transitions from the equilibrium ensemble perspective practically does not depend on this initial, short-ranged normal diffusion regime: See in part (f) of Fig. \ref{Fig2}! Hence, it compels to study also a purely sub-Ohmic subdiffusive case with $\eta_0=0$. Such a study reveals, however, in Fig. \ref{Fig3}, f that the corresponding expression in Eq. (\ref{subOhm_init}) fails badly to describe the statistics on the relevant intermediate time scale. Nevertheless, it nicely describes the initial stretched exponential kinetics with the exponent $1-\alpha/2$. Indeed, the analytical result in Eq. (\ref{Gamma_new}) yields $\Gamma\approx 121.205$, whereas the numerics imply $\Gamma\approx 112.35$, see in part (f). Here, the discrepancy is less than about 7.3\% only. In this part, our results confirm the results by Tang and Marcus \cite{TangMarcusPRL} for the residence time distribution of the initial times.  However, their prediction of the intermediate power law, $\psi(\tau)\propto 1/\tau^{2-\alpha/2}$, which also follows from our Eq. (\ref{subOhm_ML}), turns out to be wrong. The numerics are more consistent with the intermediate $\psi(\tau)\propto 1/\tau^{1+\alpha}$, whereas the tail of the distribution is again a stretched-exponential with the power exponent $\alpha$. Generalized Zusman theory fails to describe these features observed in the numerical experiments.

\section{Where and why the ensemble non-Markovian theory fails}

As we see, this theory nicely describes the relaxation of electronic populations and the initial statistics of residence time distributions of single trajectories. Moreover, it also correctly predicts that the mean residence time in the electronic states, from a single particle perspective, is always given by the inverse MLD rate, even in a profoundly adiabatic regime, even for infinitely ranged memory effects in the dynamics of the reaction coordinate. This striking feature is probably the deepest expression of a profound breaking of ergodicity in the adiabatic ET due to quantum effects. However, it badly fails to describe (i) the survival probabilities from the ensemble perspective, (ii) the tail of the residence time distribution in the case of single trajectories, and (iii) an intermediate power law regime in the case of a strictly subdiffusive dynamics on the level of single trajectories. It naturally provokes the question: Where and why the pertinent theory fails?

To answer this important question it is natural to use the picture of a multi-dimensional Markovian embedding  utilized to simulate the single trajectories in this paper. Indeed, within the Markovian embedding scheme the Eq. (\ref{basic2}) must be replaced by 
\begin{eqnarray}\label{basic_embedding}
&&p_i(x, \vec{y},t)=G_i(x,\vec{y},t|x',\vec{y'})\\
&&-v_0\int_0^tdt'\int_{-\infty}^{\infty}d\vec{y'}G_i(x,\vec{y},t-t'|x^*,\vec{y'})p_i(x^*,\vec{y'},t'),\nonumber
\end{eqnarray}
where $G_i(x,\vec{y},t|x',\vec{y'})$ is the Green function of the corresponding multi-dimensional Markovian Fokker-Planck equation. Its explicit form is not required to understand our argumentation. The 
Laplace-transformed Eq. (\ref{basic_embedding}) reads
\begin{eqnarray}\label{basic2_embedding}
&&\tilde p_i(x, \vec{y},s)=\tilde G_i(x,\vec{y},s|x',\vec{y'})\\
&&-v_0\int_{-\infty}^{\infty}d\vec{y'}\tilde G_i(x,\vec{y},s|x^*,\vec{y'})\tilde p_i(x^*,\vec{y'},s)\;.\nonumber
\end{eqnarray}
However, it is difficult to solve without further approximations for $\tilde p(x^*,s)=\int \tilde p(x^*,\vec{y}, s)d \vec{y} $. One can use e.g. the Wilemski and Fixman approximation
\begin{eqnarray}\label{Wilemski}
\tilde p(x^*,\vec{y}, s)\approx \tilde p(x^*,s)p_{\rm st}(\vec{y}),
\end{eqnarray}
where $p_{\rm st}(\vec{y})$ is the stationary distribution of the auxiliary variables. In this case,
upon introduction of the reduced propagator 
\begin{eqnarray}\label{Green_red}
 \tilde G^{(\rm red)}_i(x,s|x^*)&& =\nonumber \\
&&\int \int \tilde G_i(x,\vec{y},s|x^*,\vec{y'})p_{\rm st}(\vec{y'}) d\vec{y'} d\vec{y}
\end{eqnarray}
one can see that the problem is reduced to the previous one with $ \tilde G^{(\rm red)}_i(x,s|x^*)$ treated as a non-Markovian propagator. It is indeed nothing else the non-Markovian Green function (\ref{eta-s}), (\ref{Green}), with the memory kernel in (\ref{sum}), which corresponds to a multi-dimensional Markovian embedding description. The principal assumption here is a fast equilibration of the auxiliary variables leading to Eqs. (\ref{Wilemski}) and (\ref{Green_red}). However, this assumption is, strictly speaking, completely wrong for those modes $y_i$, which are slow on the time scale of electronic transitions. Here, we locate precisely the reason for the ultimate failure of the non-Markovian Zusman equations description. It is, in fact, heavily based on the Wilemski and Fixman approximation, which cannot be justified for the slow modes of the environment. This reason for failure is precisely the same as for the failure of non-Markovian Fokker-Planck equation to describe survival probabilities of classical bistable transitions \cite{GoychukPRE09, GoychukACP12}. One should wonder about why such a description sometimes nicely works, rather than about its failure, which is generally expected. Notably, the approach based on non-Markovian Fokker-Planck equation generally fails to describe statistics of single trajectories. Although, it can properly describe the most probable value of the logarithmically transformed residence times, in the case of classical bistable transitions \cite{GoychukPRE09, GoychukACP12}, and the mean residence time, in the present case. Moreover, in the present case, it does describe the initial part of the residence time distribution properly. However, it completely fails to describe the escape kinetics with the absorbing boundary condition at the crossing point, on the ensemble level. The reason is clear: Each electron makes a transition at a fixed, non-equilibrium and quasi-frozen realization of the reaction coordinate, whereas non-Markovian Zusman equations \textit{implicitly} assume that all the environmental modes $y_i$, which are responsible for the memory effects, are instantly equilibrated. Only, in this case, one can exclude the dynamics of $y_i(t)$ and introduce an NMFPE with Green function (\ref{Green}). However, if the \text{same} electron makes huge many transitions, in the long run, it samples different random realizations of the reaction coordinate at each transition. Then, the problem becomes essentially softened, and the description becomes well justified, on the level of population relaxation. However, it must be used with great care, when applied to single trajectories. For example, it predicts completely wrong asymptotics of the survival probabilities, and the prediction of the correct intermediate asymptotics in the case of finite $\eta_0$ is just due to Markovian character of the reaction coordinate dynamics on the corresponding time scale. However, once again, when huge many particles repeatedly jump between the electronic states this kind of non-Markovian description becomes utterly correct for the population relaxation. Most theories of electron transfer focus namely on the population relaxation, which can be, however, quite misleading and inappropriate to describe ET statistics in slowly fluctuating environments as this work shows.  

\section{Summary and Conclusions}

In this work, we elucidated the essential features of fractional electron transfer kinetics in a Cole-Cole, subdiffusive sub-Ohmic environment both from the ensemble perspective of non-Markovian Zusman equations within the contact approximation (a truly minimal semi-classical setting) and from the perspective of single trajectories, within a closely related stochastic trajectory description. Our both analytical and numerical study revealed that: 

i) In a profoundly nonadiabatic ET regime, for very small tunnel couplings, the ET kinetics viewed from the perspective of survival probabilities remains ergodic even in such slowly fluctuating environments. It is exponential and well described by the Marcus-Levich-Dogonadze rate. However, at odds with this remarkable fact, the relaxation of electronic populations to equilibrium has a universal power-law tail whose weight diminishes with diminishing electronic coupling. The smaller the tunnel coupling, the later starts this residual anomalous behavior. It can be buried in noise, and hence very difficult to reveal.

(ii) The ensemble theory based on the generalized Zusman equations remarkably well predicts the relaxation of electronic populations in the whole range of permitted $V_{\rm tun}$ variations. Our analytical result agrees very well with stochastic trajectory simulations. In the adiabatic regime, electronic relaxation is initially stretched exponential and then changes over into a power law. For some parameters, it is described by the same Mittag-Leffler functional dependence, which also describes the relaxation of the reaction coordinate. It corresponds to the Cole-Cole dielectric response, often measured in protein systems. However, the relaxation time parameter entering this electronic relaxation (and the related Cole-Cole response) is very different from one of the reaction coordinate. Interestingly enough, it depends not exponentially on the height of the activation barrier, what one generally expects (i.e., an Arrhenius dependence), but in a power law manner.

(iii) With increasing tunnel coupling, a profound violation of the kinetic ergodicity is demonstrated. Survival probabilities in electronic states start to display two very different, conflicting kinetics from the ensemble and single trajectory perspectives. This violation of ergodicity occurs both on account of long-lasting memory effects in the viscoelastic environment, and due to a profound quantum nature of electron transfer on the level of single particles, even in a seemingly classical, from the ensemble point of view, adiabatic regime.

(iv) The equilibrium ensemble theory based on the generalized Zusman equations turns out to be completely wrong in predicting the kinetic behavior of the ensemble of the particles making the transition to another state without return, and we explained the reason why. The corresponding theory predicts that the residence time distribution does not possess a mean time and has a power law tail, $\psi_i(t)\propto t^{-1-\alpha}$. Both predictions are entirely wrong. Not only the mean time but also the variance are finite, and the tail is stretched exponential. The reason for non-Markovian theory failure is that the slow viscoelastic modes of the medium are quasi-frozen and not equilibrated,  when the electron jumps out of the state at the curve-crossing point, at odds with implicit theoretical assumptions.

(v) The non-equilibrium ensemble theory applied to describe statistics of stationary, equilibrium single electron transitions correctly predicts the mean residence time even in a profoundly adiabatic regime. It is given by the inverse of MLD rate, for any medium. However, its prediction that the variance diverges in the Cole-Cole, or sub-Ohmic medium is wrong. The theory predicts that the tail of the distribution is a power law, $\psi(t)\propto t^{-2-\alpha}$. This prediction is also wrong: the tail is always stretched exponential. The theory works well in the deeply non-adiabatic regime. Also in the deeply adiabatic regime, for $\eta_0\neq 0$, it describes $90+\%$ of the initial decay of survival probability. However, this remarkable success is just because on the corresponding time scale the normal diffusion dominates, and the related analytical result corresponds to the result of Markovian theory in Ref. \cite{PCCP17}. For the strictly sub-Ohmic case of $\eta_0=0$, the theory describes very well the initial stretched-exponential decay with the power exponent $1-\alpha/2$. It is also a remarkable success. However, the intermediate power law, which the theory predicts, $\psi(t)\propto t^{-2+\alpha/2}$ does not exist, for $0<\alpha<1$. It presents an artifact of the theory based on generalized Zusman equations. This prediction, which is central for the Tang-Marcus theory of quantum dots blinking \cite{TangMarcusPRL} in non-Debye media,  is wrong.

To develop a flawless analytic theory of non-ergodic single electron transport provides a real current challenge for the theorists. Indeed, the theory based on the generalized Zusman equations can profoundly fail in some fundamental, key aspects, as our study manifested. However, the developed stochastic numerical approach to the underlying curve-crossing problem can be used reliably instead, within the same parameter range of the overall model validity. It is restricted, however, to a series of approximations, primarily to the contact approximation. To go beyond it, e.g., in the spirit of our earlier work \cite{GoychukChemPhys01}, generalized towards non-Markovian dynamics of the reaction coordinate, provides one of the exciting directions to explore in the future. The problem is, however, much more challenging and profound. Indeed, what to do in the case of a fully quantum description? The most successful current quantum theories of electron transfer are the ensemble theories based on the concept of the reduced density matrix. Our work shows that the related ensemble approach (in a semi-classical limit) fails overall to describe the statistics of single electron transitions in an adiabatic regime in the case of non-Debye media featured, e.g. by the Cole-Cole response. This \textit{inter alia} can be a common situation in the case of biological electron transfer. Most experiments, which were done thus far, present ensemble measurements, and experiments with single molecules are capable of surprises. Such experiments are highly welcome and appreciated. To develop a proper fully quantum theory based on the trajectory description also provides a real challenge, which the readers are invited to address. 

\section*{Acknowledgment} 
Funding of this research by the Deutsche Forschungsgemeinschaft (German Research Foundation), Grant GO 2052/3-1 is gratefully acknowledged.

\appendix

\section{Adiabatic time-functions}\label{appendix}

\begin{figure}
\vspace{1cm}
\resizebox{0.87\columnwidth}{!}{\includegraphics{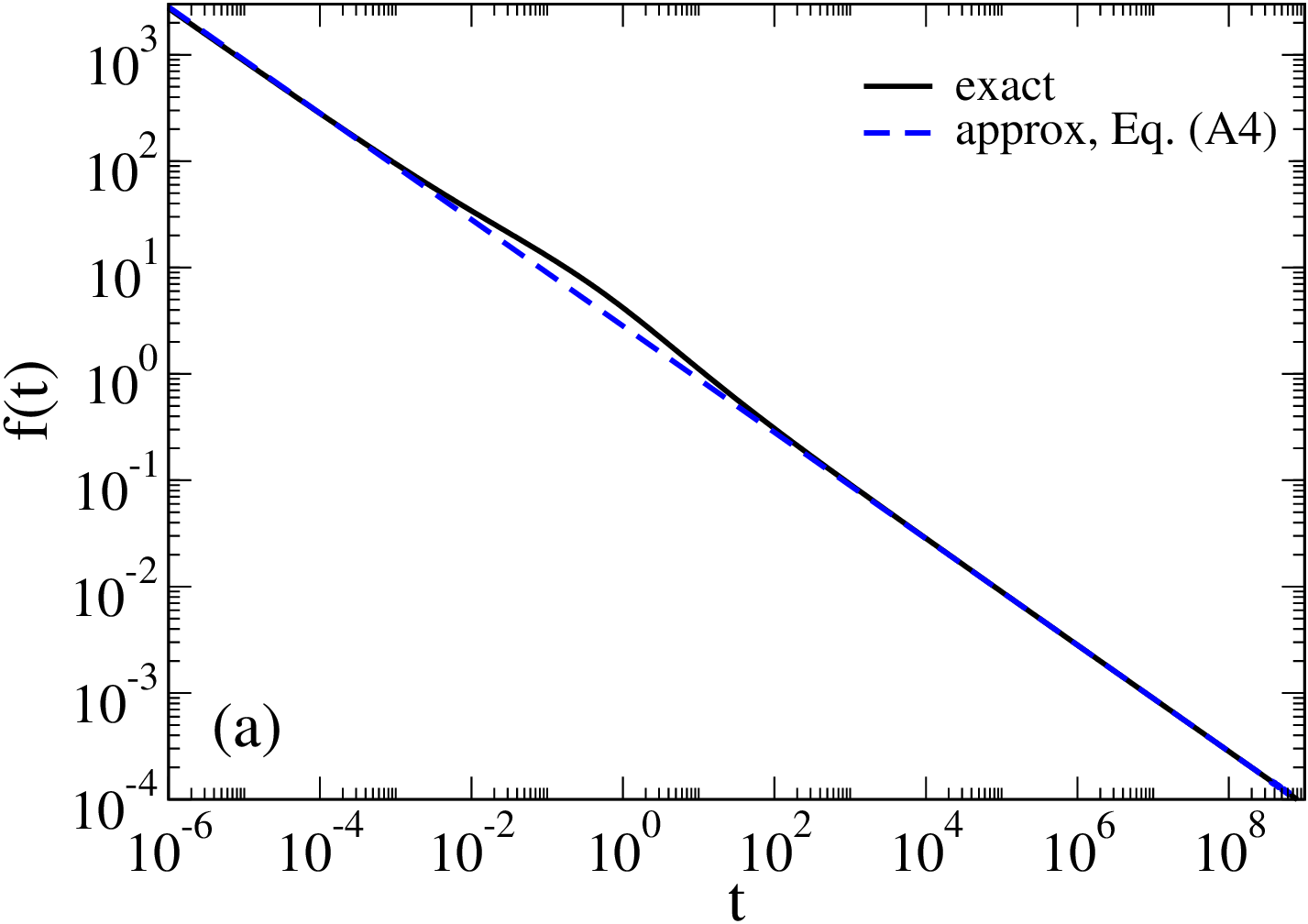}}\\[0.8cm]
\resizebox{0.93\columnwidth}{!}{\includegraphics{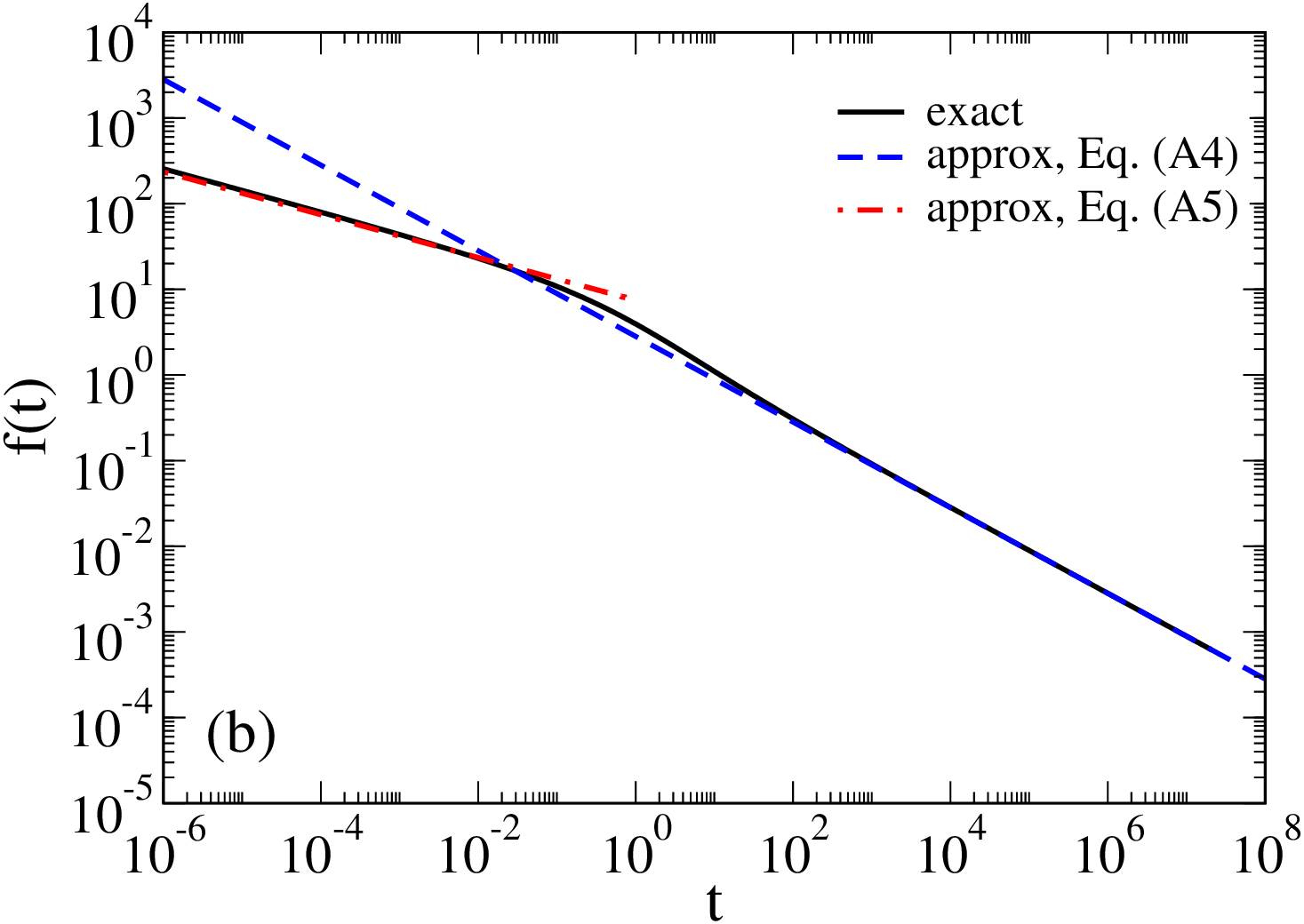}}
\caption{(color online) The sum $f(t)=f^{(1)}(t)+f^{(2)}(t)$ in Eqs. (\ref{f1}), (\ref{f2}) (full black line), and its corresponding approximations by Eq. (\ref{reg2}) (dashed blue line) or/and Eq. (\ref{reg1a}) (dash-dotted red line)
for the case $\alpha=0.5$, $r_1=r_2=2.5$ and (a) $\tilde \eta_0=0.1$ or (b) $\tilde \eta_0=0$.  Time is in units of $\tau_r$.}
\label{FigA1}       
\end{figure}

This Appendix deals with functions $\tilde \tau_{1,2}(s)$ in Eq. (\ref{tau(s)}), which have a rather complex structure and are not easy to analyze. They can be expressed as sums of two contributions, $\tilde \tau_{1,2}(s)=\tilde \tau^{(1)}(s)+\tilde \tau_{1,2}^{(2)}(s)$, where $\tilde \tau^{(1)}(s)$ is the Laplace-transform of 
\begin{eqnarray}\label{f1}
f^{(1)}(t)=\frac{1}{\sqrt{1-\theta^2(t)}}-1
\end{eqnarray}
and $\tilde \tau^{(2)}_{1,2}(s)$ is the Laplace-transform of 
\begin{eqnarray}\label{f2}
f_{1,2}^{(2)}(t)=\frac{1}{\sqrt{1-\theta^2(t)}}\left ( e^{2r_{1,2}\theta(t)/[1+\theta(t)]} -1 \right ),
\end{eqnarray}
where $\theta(t)$ is the coordinate relaxation function and $r_{1,2}=E_{1,2}^{(a)}=(\lambda \mp \epsilon_0)^2/(4\lambda k_BT)$ are activation energies of ET in the units of $k_BT$. We restrict our analysis to an important parameter regime of sufficiently large activation barriers $r_{1,2}\gtrapprox 2$. Then, the first contribution in the sum can be neglected and we concentrate on the function $f^{(2)}(t)$, where we drop subindex for a while. 
We are interested in the case $z=\tau_0/\tau_r\ll 1$, where the relaxation of the reaction coordinate can be approximately described by (\ref{theta_approx}), except for the initial times
$t<z\tau_r$. Notice, that the scaled $\tilde \eta_0=\eta_0/(\eta_\alpha\tau_r^{1-\alpha})=z$.
Next, we consider two parameter regimes: (i) $t\ll \tilde \eta_0\tau_r$, (ii) $t\gg \tau_r$. In the first one, $\theta(t)\approx \exp[-t/(z\tau_r)]\approx 1- t/(z\tau_r)$, and we have
\begin{eqnarray}\label{reg1}
f^{(2)}_{1,2}(t)\approx  \frac{c_1}{\sqrt{t/\tau_r}} 
\end{eqnarray}
with $c_1=\sqrt{\frac{\tilde \eta_0}{2}}\left (e^{r_{1,2}}-1\right)$ universally for any $\alpha$. By an Abelian theorem \cite{Doetsch} this yields (\ref{approx2}).
In the second regime, $\theta(t)\sim (1/\Gamma(1-\alpha))(t/\tau_r)^\alpha\ll 1$, and we have
\begin{eqnarray}\label{reg2}
f^{(2)}_{1,2}(t)\approx  \frac{c_2}{(t/\tau_r)^\alpha} 
\end{eqnarray}
with $c_2=2r_{1,2}/\Gamma(1-\alpha)$.  By a Tauberian theorem \cite{Doetsch} this yields (\ref{approx}). In the numerical studies of this paper, we consider a symmetric ET with $\alpha=0.5$, $r_1=r_2=2.5$ and $\tilde \eta_0=0.1$. In this particular case, $c_1\approx 2.50$ and $c_2\approx 2.82$. This is the reason why the approximation (\ref{approx}) works well in the whole range of the variable $s$, see in Fig. \ref{FigA1}, a. This is, however, a lucky case beyond which the beauty of the related analytical results in the adiabatic ET regime is lost. Generally,  short and long time asymptotics in (\ref{reg1}) and (\ref{reg2}) are very different even for $\alpha=0.5$, since $c_1$ and $c_2$ can differ strongly, in general. For $\alpha=0.5$,  one must approximately satisfy $c_1\approx c_2$, or $\sqrt{\frac{\tilde \eta_0}{2}}\left (e^{r}-1\right)\approx 2r/\sqrt{\pi}$, for the approximation (\ref{approx}) to work uniformly.  This can  be done only in a symmetric case.

Furthermore, for a model with $\eta_0=0$ (strictly sub-Ohmic environment), Eq. (\ref{reg1}) is replaced for $t\ll \tau_r$ by
\begin{eqnarray}\label{reg1a}
f^{(2)}_{1,2}(t)\approx  \frac{c_3}{(t/\tau_r)^{\alpha/2}} 
\end{eqnarray}
with $c_3=\sqrt{\Gamma(1+\alpha)/2}\left (e^{r_{1,2}}-1\right)$. 
This asymptotics yields (\ref{approx3}) for $s\tau_r\gg 1$. 
Notice that in this case, the power-law behaviors for $t\ll \tau_r$ and $t\gg \tau_r$ are very different, see in Fig. \ref{FigA1}, b. 

\bibliographystyle{apsrev4-1}
\bibliography{LZfin}

\begin{thebibliography}{132}
\bibitem [{foo({\natexlab{a}})}]{footnote}%
  \BibitemOpen
  \href@noop {} {}  \bibinfo {note} {Surprisingly, Debye
  relaxation constant of biological bound water can reach even nanoseconds
  \cite{Nandi18}.}\BibitemShut {Stop}%
\bibitem [{foo({\natexlab{c}})}]{footnote2}%
  \BibitemOpen
  \href@noop {} {}  \bibinfo {note} {Eq. (\ref{Zusman})
  should be distinguished from the equations of Sumi-Marcus \cite{Sumi86} or
  Agmon-Hopfield \cite{Agmon84} models generalized to the case of non-Debye
  environments \cite{Zhu92}, which look similar for the case of symmetric
  electron transfer, $\epsilon_0=0$. Similar approaches have also been
  developed within a stochastically driven spin-boson model
  \cite{GoychukPRE95,GoychukJCP95,GoychukPRE97,GoychukAP05}. Within the
  generalized Sumi-Marcus model, $K(x)$ is a non-adiabatic ET rate, like one
  below in Eq. (\ref{na-rate}). It is modulated by a slow conformational degree
  of freedom $x$, via e.g. modulation of the energy bias,
  $\epsilon_0\to\epsilon_0(x)$. The generalized Sumi-Marcus approach
  corresponds to a slow conformational dynamics modulating non-adiabatic
  electron transfer via the conformation-dependent activation energies
  $E_{1,2}^{(a)}(x)$. Likewise, slow conformational dynamics can affect the
  effective tunnel coupling \cite{Yang03,GoychukJCP05} leading also to a slowly
  fluctuating non-adiabatic rate. However, in the so-called narrow reaction
  window limit $K(x)$ is approximated by a delta-function in the Sumi-Marcus
  theory. This relation makes our theory and results of much broader importance
  in the context of long-range electron transfer. However, in this work we
  interprete the results within a generalized Zusman theory. In the considered
  case, $E_{1,2}^{(a)}$ and $V_{\rm tun}$ are constant.}\BibitemShut {Stop}%
\end{thebibliography}

\end{document}